\documentclass[preprint,12pt]{elsarticle}

\usepackage{amsmath,amsthm,amsfonts,amssymb,amscd}
\usepackage{graphicx}
\usepackage{algorithm}
\usepackage{algpseudocode}

\usepackage[utf8]{inputenc}
\usepackage{fullpage,xcolor}
\usepackage{cite}
\usepackage{mathtools}

\usepackage{subcaption}
\usepackage{caption}
\usepackage[breaklinks=true, colorlinks=true, linkcolor={blue!90!black}, citecolor={blue!90!black}, urlcolor={blue!90!black}]{hyperref}

\graphicspath{{}}

\begin{document}
\begin{frontmatter}

\title{Implementation of a Mesh refinement algorithm into the quasi-static PIC code QuickPIC}

\author[ucla]{Q. Su}
\ead{xpsqq@g.ucla.edu}
\author[tsinghua]{F. Li}

\author[normal,normal2]{W. An}

\author[uclap]{V. Decyk}

\author[uclap,ucla]{Y. Zhao}

\author[uclap,ucla]{L. Hildebrand}

\author[uclap,ucla]{T. N. Dalichaouch}

\author[tsinghua]{S. Zhou}

\author[uclap,uclat]{E. P. Alves}

\author[lbnl]{A. S. Almgren}

\author[ucla,uclap]{W. B. Mori}

\affiliation[ucla]{organization={Department of Electrical and Computer Engineering, University of California},
            city={Los Angeles},
            postcode={90095}, 
            state={CA},
            country={USA}}
\affiliation[uclap]{organization={Department of Physics and Astronomy, University of California},
            city={Los Angeles},
            postcode={90095}, 
            state={CA},
            country={USA}}   

\affiliation[uclat]{organization={Mani L. Bhaumik Institute for Theoretical Physics, University of California},
            city={Los Angeles},
            postcode={90095}, 
            state={CA},
            country={USA}}            

\affiliation[tsinghua]{organization={Department of Engineering Physics, Tsinghua University},
            city={Beijing},
            postcode={100084}, 
            country={China}}            

\affiliation[normal]{organization={Institute for Frontiers in Astronomy and Astrophysics, Beijing Normal University},
            city={Beijing},
            postcode={102206}, 
            country={China}}

\affiliation[normal2]{organization={Department of Astronomy, Beijing Normal University},
            city={Beijing},
            postcode={100875}, 
            country={China}}             
\affiliation[lbnl]{organization={Lawrence Berkeley National Laboratory},
            city={Berkeley},
            postcode={94720}, 
            country={USA}}             
\begin{abstract}
 Plasma-based acceleration (PBA) has emerged as a promising candidate for the accelerator technology used to build a future linear collider and/or an advanced light source. In PBA, a trailing or witness particle beam is accelerated in the plasma wave wakefield (WF) created by a laser or particle beam driver. The WF is often nonlinear and involves the crossing of plasma particle trajectories in real space and thus particle-in-cell methods are used. The distance over which the drive beam evolves is several orders of magnitude larger than the wake wavelength. This large disparity in length scales is amenable to the quasi-static approach. Three-dimensional (3D), quasi-static (QS), particle-in-cell (PIC) codes, e.g., QuickPIC, have been shown to provide high fidelity simulation capability with 2-4 orders of magnitude speedup over 3D fully explicit PIC codes. In PBA, the witness beam needs to be matched to the focusing forces of the WF to reduce the emittance growth. In some linear collider designs, the matched spot size of the witness beam can be 2 to 3 orders of magnitude smaller than the spot size (and wavelength) of the wakefield. Such an additional disparity in length scales is ideal for mesh refinement where the WF within the witness beam is described on a finer mesh than the rest of the WF. We describe a mesh refinement scheme that has been implemented into the 3D QS PIC code, QuickPIC. We use a very fine (high) resolution in a small spatial region that includes the witness beam and progressively coarser resolutions in the rest of the simulation domain. A fast multigrid Poisson solver has been implemented for the field solve on the refined meshes and a Fast Fourier Transform (FFT) based Poisson solver is used for the coarse mesh. The code has been parallelized with both MPI and OpenMP, and the parallel scalability has also been improved by using pipelining. A preliminary adaptive mesh refinement technique is described to optimize the computational time for simulations with an evolving witness beam size. Several test problems are used to verify that the mesh refinement algorithm provides accurate results. The results are also compared to highly resolved simulations with near azimuthal symmetry using a new hybrid QS PIC code QPAD that uses a PIC description in the coordinates ($r$, $ct-z$) and a gridless description in the azimuthal angle, $\phi$. 

 \end{abstract}

 \end{frontmatter}
\section{Introduction}	

The study of interactions between intense lasers or beams and particles has become a vibrant and rapidly developing area of plasma and beam physics research. One of the applications of this field of research is the development of plasma-based accelerators (PBA)~\citep{joshi2003plasma, 10.1063/5.0004039}, which can be used for next generation advanced light sources and a future linear collider. In PBA, an intense laser (LWFA)~\citep{PhysRevLett.43.267} or a particle beam (PWFA)~\citep{PhysRevLett.54.693} is sent through a tenuous plasma. The radiation pressure of the laser or the space charge fields of a particle beam can create a plasma wave wakefield (WF) in PBA, which has a phase velocity that is close to the speed of light. There have been many successful experimental results in PWFA and LWFA~\citep{blumenfeld2007energy, litos2014high, gonsalves2011tunable, kneip2010bright, adli2018acceleration, corde2015multi, nie2018relativistic}, demonstrating the potential of PBAs for various applications.

The success of these experiments has benefited from advanced plasma simulation tools. The standard electromagnetic (EM) particle-in-cell (PIC) methods~\citep{birdsall2018plasma} have been developed to study PWFA and LWFA. The standard PIC code advances the fields forward in time by solving the full set of temporally and spatially discretized Maxwell's equations. The current and charge density in these equations is computed by summing the contributions from each particle. Each particle contributes a fraction of its current and charge to each grid point based on a chosen deposition order. The updated fields on the grid are then interpolated at the particle locations. These interpolated fields are then used to update the particle's momentum and position. Depending on the type of Maxwell solver the fields and currents/charge densities are defined on the grid or within a cell. These full EM PIC codes have been very successful at studying a variety of problems in plasma physics, including PBA, where details on the trajectories of individual particles are required. However, the computational expenses of the full 3D PIC algorithms can be formidable for some applications due to the Courant-Friedrichs-Lewy (CFL) condition~\citep{courant1928partiellen}, which sets a limit on the time step size.

In PBA, the duration (length) of the driver and the period (wavelength) of the wakefield are comparable while the propagation distance is more than three orders of magnitude larger. The quasi-static approximation (QSA) \citep{sprangle1990nonlinear} was proposed  to exploit this disparity of scales for efficient simulations of intense laser or beam plasma interactions. The QSA takes advantage of the difference in time scales between the high-energy beam or laser and the plasma wake evolution. The QSA uses variables $(x, y, \xi=c t-z; s=z)$ to describe the field equations and plasma dynamics instead of $(x, y, z; t)$, where c is the speed of light. The evolution of all quantities can be separated into two time scales, one for $\xi$ and the other for $s$.

In the QSA, the driver is assumed to not evolve, i.e., it is static, during the time it takes to pass by a plasma particle. Thus, plasma electrons are assumed to have longitudinal speeds $v_z$ such that 
$(v_b-v_z)/c>\epsilon$ where $v_b$ is the beam velocity and $\epsilon$ is a smallness parameter that scales $1/(k_\beta c\tau)$, where $\tau$ is the driver pulse length and $k_\beta \equiv k_p / \sqrt{2 \gamma} $ is the betatron wavelength of the beam (for a laser $v_b$ is the group velocity of light in plasma and $k_\beta^{-1}$ is replaced with the Rayleigh length), where $k_p \equiv \omega_p / c$, and $\omega_p$ is the plasma frequency. Under these conditions, $\partial_{s}$ for plasma quantities (the change with the lab frame distance along the beam direction) is much smaller than $\partial_{\xi}$ (the change with the position in the co-moving window), which allows for a simplified description of the system. A quasi-static PIC code can be developed by assuming the driver quantities (e.g. laser or beam) are fixed when solving for the plasma related quantities. After solving for the plasma quantities, the fields can then be used to advance the beam particles for a single ``large" time step. The updated beam particles are then utilized to solve for plasma quantities in the next 3D step. Several PIC codes have been developed based on the QSA, such as WAKE~\citep{mora1997kinetic}, LCODE~\citep{lotov2003fine}, QuickPIC~\citep{huang2006quickpic,an2013improved}, HiPACE$++$~\citep{mehrling2014hipace, diederichs2022hipace++},  INF\&RNO \citep{benedetti2012efficient},
WANDPIC~\citep{wang2020wand}, and QPAD ~\citep{li2021quasi}. 
WAKE, LCODE, and INF\&RNO are 2D (r-z geometry) quasi-static PIC codes, while QuickPIC, HiPACE++, and WANDPIC are 3D. QPAD is a quasi-3D QSA based PIC code where the fields are expanded into a truncated series of azimuthal harmonics. QuickPIC is a 3D QS PIC code that has been fully parallelized (MPI+OpenMP with pipelining). It has been shown to achieve speedups of $10^2 \sim 10^4$ without loss in accuracy compared with a full PIC code (e.g. OSIRIS \citep{fonseca2002osiris}).

Despite the high speedup compared to a full PIC code, quasi-static PIC codes may still incur high computational expenses in situations where there is a significant discrepancy in scale between the transverse dimensions of the acceleration structure and the witness beams. For example, in the design of a plasma wakefield acceleration  (PWFA) based linear collider for electron acceleration~\citep{adli2013beam}, a relativistic drive beam excites a plasma wake, with the trailing beam placed in the accelerating phase of the wake. In future linear collider designs, the accelerated beam must meet certain requirements, one of which is high luminosity. The luminosity is calculated as $L=\left(f N^2 / 4 \pi \sigma_x \sigma_y\right)$, where $N$ represents the number of particles in each bunch, $f$ is the frequency of the collisions, and $\sigma_x, \sigma_y$ is the spot size of the bunch at the interaction point. To achieve or exceed the luminosity goal of a linear collider, $L \sim 10^{34} \mathrm{~cm}^{-2} \mathrm{~s}^{-1}$, a witness beam must be focused to a small spot size at the interaction or collision point. This necessitates that the beam have a normalized emittance of $\sim 100 \mathrm{~nm}$ and  charge of $\sim 1~\mathrm{nC}$. As the electron and positron beams collide they pinch which can lead to large amounts of radiation. The use of asymmetric beams with the same RMS spot size can reduce this beamstrahlung for the same luminosity~\citep{adli2013beam}. This is achieved through the use of asymmetric emittances. In some designs, asymmetries of $\sigma_x / \sigma_y \sim 100$ are utilized~\citep{adli2013beam}. To suppress the emittance growth within the uniform plasma accelerator stage, a beam needs to have a matched spot size $\sigma_r=\left(2 \epsilon_N^2 / \gamma\right)^{1 / 4}$, where $\epsilon_N$ represents the normalized emittance of the beam and $\gamma$ denotes the relativistic Lorentz factor of the beam. For such a tightly focused high current beam, the Coulomb field will pull the ions inward ~\citep{lee1999wakefield, PhysRevLett.95.195002} within the transit time of the beam. In order to properly resolve the ion motion, the simulation must resolve the spot size of the beam and the simulation window must encompass the entire acceleration structure. For linear collider parameters, the matched spot size of the trailing beam can be hundreds of times smaller than the acceleration bubble radius. The computational demands to model the entire domain are substantial even with a 3D quasi-static code. This is challenging for contemporary supercomputers due to both memory and CPU requirements. In Ref.~\citep{an2017ion}, this was addressed by only modeling a small region near the witness beam. However, in order to model the effects of ion motion on beam loading and hosing instabilities, the entire WF region must be included. Thus, enhancing the computational efficiency of the quasi-static code is vital for linear collider simulations.

Several strategies have been proposed to address these challenges. One such approach is azimuthal Fourier decomposition ~\citep{lifschitz2009particle}. In this hybrid method, the fields and charge densities (or current densities) are decomposed into multiple azimuthal modes, and the high-order modes with low amplitude can be truncated based on the degree of asymmetry. A PIC description is used on an $r-z$ grid, and a gridless description is used for the azimuthal angle $\phi$. For example, an azimuthally symmetric beam can be fully represented by the first harmonic. This method has been applied to full PIC codes~\citep{lifschitz2009particle, davidson2015implementation, lehe2016spectral}. However, even with this approach the use of full PIC simulations is not feasible due to the 
CFL~\citep{courant1928partiellen}
limit and physics constraints on the time step for small transverse cell sizes. Recently, the azimuthal mode expansion has also been applied to the QuickPIC workflow, resulting in a new code  QPAD~\citep{li2021quasi}. This provides speedups of around $O(N)$ for problems with low azimuthal asymmetry, where $N$ is the grid numbers in one transverse direction. Another approach is mesh refinement (MR), which involves increasing the resolution in areas where a high level of detail is needed. Some mesh refinement capability has been implemented into full PIC codes~\citep{vay2018warp}, however, time step constraints based on the smallest cell sizes are still present. Mesh refinement has also been implemented into 2D ~\citep{benedetti2017efficient} and 3D  ~\citep{mehrling2018subgrid,diederichs2022hipace++} quasi-static codes. The quasi-static algorithm necessitates solving a series of 2D Poisson like equations. The basic MR idea is to solve these equations globally on a coarse grid and then use the solutions to obtain boundary conditions for solving the same equations on a fine mesh within a much smaller domain. However, there is limited published detail on the implementations and on benchmarking the existing algorithms.

In this article, we describe in much detail how a mesh refinement algorithm is implemented into the quasi-static PIC code QuickPIC. The code applies high resolution refined meshes in specific regions surrounding the trailing beam, and the grid sizes of those meshes progressively change to coarser resolutions in the remaining parts of the simulation domain. A fast multigrid Poisson solver has been implemented to solve 2D Poisson like equations on the refined meshes, and the solver has been parallelized with both MPI and OpenMP. The equations for the coarsest mesh can be solved using either Fast Fourier Transform (FFT) or multigrid algorithms. In the current implementation, we use FFTs to solve the equations on the coarsest mesh that covers the whole simulation domain. 

In order to tackle the issue of imbalanced computational workload, we distribute the Poisson solvers on the fine grids, usually concentrated in a single location, to all processors. Afterwards, we collect the results to the appropriate processor so that the particles can be pushed to new momenta and positions. The scalability of the code has also been improved significantly by using pipelining ~\citep{feng2009enhancing} in the $\xi$ direction. Additionally, the code allows for inhomogeneous particle loading, which enables the initialization of a greater number of plasma particles within the fine mesh region when necessary, while employing a lower number of particles in the coarse mesh regions. A preliminary adaptive mesh refinement technique is also implemented in order to optimize the computational time for simulations with an evolving beam size. This is achieved by modifying the refined mesh size and resolution to meet the simulation's demands. 

In this work, we do not directly address the self forces at the refinement boundary~\citep{vay2002mesh,colella2010controlling}. However, for all results shown, there is no spurious behavior at the refinement boundary, and this is partly due to two reasons. To control issues with particles moving between domains with different resolutions, we implemented a hierarchy of resolution layers between the coarsest and finest resolutions. In addition, we choose the locations of the boundaries between different layers of resolution to be in regions where there are not large amounts of particles crossing.  

Several problems of interest are chosen to test and benchmark the algorithm. The obtained results demonstrate consistency with those previously reported in published papers using QuickPIC. Furthermore, the benchmarking has benefited from the use of the quasi-3D QPAD, enabling simulations where the finest resolution is applied uniformly across the entire domain. Adaptive mesh refinement has also been used to simulate the evolution of an unmatched beam with mobile ions, and the results agree with fully resolved QPAD simulations.

\section{The implementations of mesh refinement in QuickPIC}

\subsection{The quasi-static loop with mesh refinement}
In this section, we provide details for how the quasi-static PIC loop in QuickPIC is modified to incorporate mesh refinement. 
\subsubsection{Review of quasi-static loop in QuickPIC}

We start with a review of the basic quasi-static PIC loop and equations in QuickPIC~\citep{huang2006quickpic,an2013improved}. To simplify the equations we use normalized units. Charge is normalized to electron charge $e$; density to the plasma density $n_0$; length to the plasma skin depth $k_p^{-1}$ defined as $c / \omega_{p}$, where $\omega_{p}$ is the plasma oscillation frequency and $c$ the speed of light; charge density $\rho$ to $e n_{p}$; current density $\vec{J}$ to $e n_{p} c$; particle momentum to $m c$; electric and magnetic fields  $(\vec{E},\vec{B})$ to $m c \omega_{p} / e$; and potentials $(\phi, \vec{A})$ to $m c^{2} / e$. In the physical examples used throughout this article, we will assume a plasma density and then the translation from normalized units to physical units can be calculated accordingly. 

\begin{figure}[!h]
	\centering
	\includegraphics[width=0.5\linewidth]{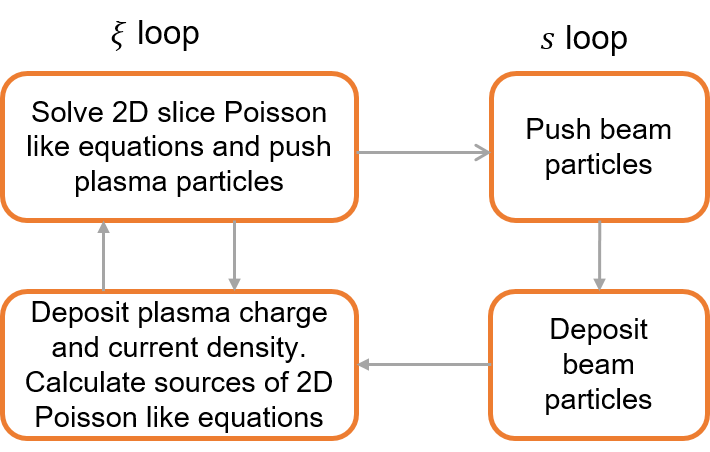}
	\caption{The quasi-static loop in QuickPIC is comprised of two primary nested loops: The inner loop in $\xi$ involves 2D field solves and plasma particle pushing, and the outer loop in $s$ is associated with beam particle operations. In the $\xi$ loop, the sources of the 2D equations (Eq. \ref{eq:poisson}) consist of operations on the deposited plasma particles and beam particles. The plasma particles are advanced in $\xi$ for a fixed $s$ based on the 2D field solutions. The outer loop consists of advancing the beam particles forward in $s$ using the plasma quantities and fields for all $\xi$ at a fixed $s$ obtained in the inner loop. Subsequently, the beam particles are deposited onto the 3D grid for use in the next $\xi$ loop, and the process repeats itself in the next loop iteration. In mesh refinement algorithm, each operation within the $\xi$ and $s$ loops is performed on both refined and coarse meshes.
}\label{fig:flow}
\end{figure}

The quasi-static equations are derived by making a mathematical transformation from $(x, y, z, t)$ to the co-moving frame variables $(x, y, \xi=c t-z, s=z)$. The partial derivatives with respect to $t$ and $z$ are then transformed to $\partial_{z}=\partial_{s}-\partial_{\xi}$ and $\partial_{t}=c \partial_{\xi}$. Mathematically, the quasi-static approximation assumes that $\partial_s$ is much smaller than $\partial_\xi$ for the fields. Thus, the evolution of all plasma quantities in PBA simulations can be separated into a fast time (spatial) scale $\xi$ and a slower scale $s$. The workflow of the QuickPIC quasi-static loop is shown in Fig.~\ref{fig:flow}. By applying quasi-static approximation $\partial_{s} \ll \partial_{\xi}$, the wave equations in the Lorenz gauge can be simplified into a set of 2D Poisson like equations \citep{an2013improved}:

 \begin{equation}\label{eq:poisson}
\begin{array}{l}
{\nabla_{\perp}^{2} \psi=-\left(\rho-J_{z}\right)},\\
{\nabla_{\perp}^{2} \vec{B}_{\perp}=\hat{z} \times\left(\frac{\partial}{\partial \xi} \vec{J}_{\perp}+\nabla_{\perp} J_{z}\right)}, \\
{\nabla_{\perp}^{2} B_{z}=-\nabla_{\perp} \times \vec{J}_{\perp}}, \\
{\nabla_{\perp}^{2} \vec{E}_{\perp}=\nabla_{\perp} \rho+\frac{\partial}{\partial \xi} \vec{J}_{\perp}},
\\ {\nabla_{\perp}^{2} E_{z}=\nabla_{\perp} \cdot \vec{J}_{\perp}},\end{array}
 \end{equation}

where $\nabla_{\perp} \equiv \hat x \partial_x+\hat y \partial_y$, $\psi \equiv \phi - A_z$ is the wake potential. QuickPIC uses FFTs to solve these equations on a uniform grid with equidistant points. The equation for $\nabla_{\perp}^{2} \vec{E}_{\perp}$ is not directly solved in QuickPIC. Instead, the transverse electric field  $\vec E_{\perp}$ is obtained indirectly through the gradient of the scalar potential $\psi$ and the transverse magnetic field $B_{\perp}$, $\vec{E}_{\perp}+\hat z \times \vec {B}_{\perp}=-\nabla_{\perp} \psi$. The source terms used in Eq.~\ref{eq:poisson} are obtained from operations on plasma density and current deposited from 2D plasma particles, and beam density deposited from 3D beam particles. Details can be found in Ref.~\citep{an2013improved}. 

At a fixed $s$, the plasma particles move in a 5 dimensional (5D) phase space, $x,y,p_x,p_y,p_z$ where $\xi$ is a time like variable. This phase space is updated every 2D step. The transverse momentum of plasma particles $\vec{P}_{\mathrm{p} \perp}$ are advanced through the equation of motion,

 \begin{equation}\label{eq:plasma_push}
\frac{\mathrm{d} \vec{P}_{\mathrm{p} \perp}}{\mathrm{d} \xi}=\frac{q_{\mathrm{p}}}{1-V_{\mathrm{p} z}}\left[\vec{E}_{\perp}+\left(\vec{V}_{\mathrm{p}} \times \vec{B}\right)_{\perp}\right], 
\end{equation}

where $\vec{V}_{\mathrm{p}}$ is the velocity of plasma particles and $q_p$ represents the normalized charge of plasma particles. The momentum of plasma particles in the z direction $P_{\mathrm{p} z}$ is derived from the conservation equation $\gamma_{\mathrm{p}}-P_{\mathrm{p} z}=1-q_{\mathrm{p}} \psi$ , leading to 

\begin{equation}\label{eq:plasma_pz}
P_{\mathrm{p} z}=\frac{1+P_{\mathrm{p} \perp}^{2} - \left[1-q_{\mathrm{p}} \psi \right]^{2}}{2\left[1-q_{\mathrm{p}} \psi \right]}.
\end{equation}

The plasma particles are then pushed to their new positions using,

\begin{equation}\label{eq:plasma_push_x}
\frac{\mathrm{d} \vec x_{\mathrm{p} \perp}}{\mathrm{d}
\xi}=\frac {\vec P_{\mathrm{p} \perp}}{1-q_b \psi},
\end{equation}
The 3D beam particles are advanced in time using the variable $s$, 

\begin{equation}\label{eq:beam_push}
\frac{\mathrm{d} \vec{P}_{\mathrm{b} \perp}}{\mathrm{d} s}=-q_{\mathrm{b}}\nabla_{\perp} \psi, \frac{\mathrm{d} P_{\mathrm{b} z}}{\mathrm{~d} s}=q_{\mathrm{b}} E_z.
\end{equation}

The scale difference that follows from the  quasi-static approximation, $\partial_{s} \ll \partial_{\xi}$, enables QuickPIC to use a relatively larger step for $\Delta s$ than for $\Delta \xi$, resulting in significant computational time savings compared to a full PIC code.

QuickPIC uses a predictor-corrector iteration loop as described in \citep{an2013improved}. The challenge is that the system of equations is not properly time-centered. This is due to the $\vec{J}_{\perp}$ and $\frac{\partial}{\partial \xi} \vec{J}_{\perp}$ terms in Eq.~\ref{eq:poisson}. Thus at each index for $\xi$ where the particle momenta and positions are updated, $\vec J_{\perp}$ and $\frac{\partial}{\partial \xi} \vec{J}_{\perp}$ are first predicted based on a previous index for $\xi$. During each iteration, the resulting fields from this prediction are then used to advance the plasma particles. This iterative process results in more accurate estimates for $\vec J_{\perp}$ and $\frac{\partial}{\partial \xi} \vec{J}_{\perp}$ which are then used to obtain more accurate values for the fields and particle momenta and locations. This iteration is carried out until a targeted level of accuracy is achieved. During this iteration process, the transverse magnetic field $B_{\perp}$ is updated using the following equation:

 \begin{equation}
\nabla_{\perp}^{2} \vec{B}_{\perp}^{\text {new }}-\vec{B}_{\perp}^{\text {new }}=\hat{z} \times\left(\frac{\partial}{\partial \xi} \vec{J}_{\perp}+\nabla_{\perp} \cdot J_{z}\right)-\vec{B}_{\perp}^{\text {old }}.
\end{equation}

\subsubsection{Mesh refinement implementation in QuickPIC}
The mesh refinement code inherits the general workflow of the QuickPIC quasi-static loop, as depicted in Fig.~\ref{fig:flow}. It initializes refined 2D meshes to solve equations (Eq.~\ref{eq:poisson}) progressively with higher resolution surrounding  regions where increased precision is needed. In this paper, we assume square cells in each domain. We designate each level of resolution with two variables: $l$ denotes the refinement level, while $r$ represents the logarithm base 2 of the refinement ratio between the current level $l$ and the next coarser level $l-1$. This ensures that the refinement ratios between $l$ and $l-1$ are constrained to be $2^r$. The mesh refinement calculations are conducted on a hierarchy of meshes denoted by $\Omega_{l, r}$. For instance, if there are three levels of refinement in total, with the second level of refinement having a refinement ratio of 8, and the third level of refinement having a refinement ratio of 4 to the second level, we will denote the three levels of meshes as $\Omega_{1,1}$, $\Omega_{2,3}$, and $\Omega_{3,2}$. Then the domain with the finest resolution in the simulation has grid size 32 times smaller in $x$ and $y$ than the coarsest resolution. In this paper,when information of the refinement ratios is given, we sometimes omit $r$ and represent $\Omega_{l, r}$ as $\Omega_l$.

\begin{figure}[!h]
	\centering
	\includegraphics[width=0.8\linewidth]{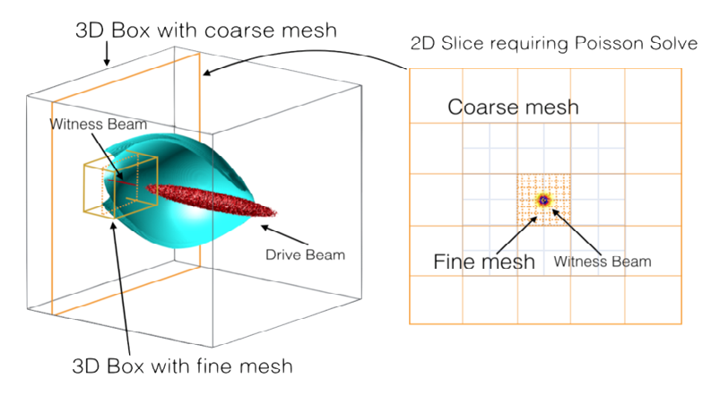}
	\caption{An example of simulating PWFA with mesh refinement. A 3D fine mesh is set up around the trailing beam with a small spot size. An intermediate level (blue grid) is employed to create a gradual transition between the fine and coarse mesh boundaries.}\label{fig:refine}
\end{figure}

A PBA example is illustrated in Fig.~\ref{fig:refine}, where an electron drive beam (red) is sent through a tenuous plasma and excites a nonlinear wake by expelling (blowing out) the plasma electrons (blue). A narrow  witness beam (red) with high current is placed behind the drive beam at an appropriate phase of the wakefield where it is accelerated and focused. As described in the introduction, the space charge field of this tightly focused witness beam
pulls ions inward where they collapse on the axis creating a very narrow density filament. The transverse dimension of this region is more than two orders of magnitude smaller than the WF width. Thus, the region near the witness beam requires much higher resolution than the rest of simulation domain. We initialize 3D refined meshes encompassing the narrow trailing beam, ensuring accuracy in the region where ion motion takes place. The hierarchical refinement system allows for multiple levels of refinement ($\Omega_{l,r}$), with each higher level containing a finer resolution mesh enclosed within a coarser mesh of the lower level. The refinement ratio between any two adjacent levels is adjustable as $2^r$.  In the example in Fig.~\ref{fig:refine}, there are three levels of resolution.

Both the field solver and particle advance,  including field interpolation and source deposit components of the quasi-static loop (Fig.~\ref{fig:flow}) have been modified to accommodate the new grid hierarchy. An FFT+multigrid solver (\ref{sec:appendix}) is implemented to replace the original FFT solver that can only solve homogeneous grids. We apply the multigrid solver to solve the 2D Poisson like equations for each refinement level of Eq. \ref{eq:poisson}, except for the coarse grid, which continues to be solved by the FFT solver.  We deposit and push plasma particles using both the fine and coarse resolution solutions, based on their respective positions. The particle operations needed for the source depositions,  (e.g. $\rho, J$), on the right hand side of Eq. \ref{eq:poisson} are performed on a hybrid grid consisting of both coarse and fine meshes.  The same approach is applied to beam particles as well. Further details regarding these modifications are elaborated below.

\begin{figure}[!h]
	\centering
	\includegraphics[width=0.5\linewidth]{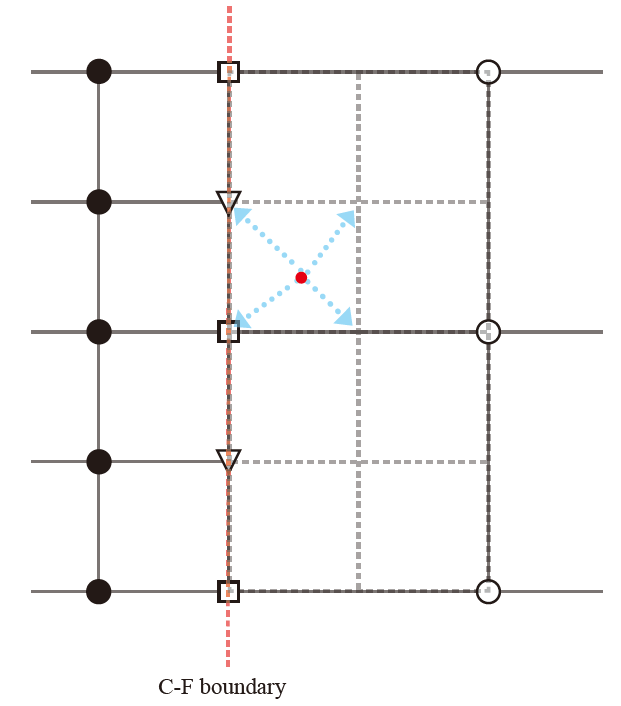}
	\caption{
Illustration of the grid configuration near a Coarse-Fine (C-F) refinement boundary. The left-hand side exhibits the fine mesh, represented by solid black dots indicating inner points. The hollow square and triangle points represents the grid points on the C-F boundary, where the square points are shared by the coarse grid. The hollow dots represent the grid that is only covered by the coarse grid. When solving Poisson like equations on the fine mesh, only the sources within the inner region of the fine mesh are considered. A single plasma particle is denoted by a red dot, and its charge or current is deposited onto the fine mesh. Particles are deposited on the fine mesh extending one coarse cell (dashed-gray grids) beyond the boundary of the region where a field quantity is being solved.}\label{fig:grid_in}
\end{figure}

For simplicity, we express and generalize these Poisson like equations in Eq.~\ref{eq:poisson} as $\nabla_{\perp}^2 F = S,  
\nabla_{\perp}^2 F - F = S$, with $F$ representing fields such as $\psi, \vec{E}$ and  $\vec{B}$.
Multiple refinement grids $\Omega_{l, r}$ identified by index $l$ and $r$ are created with grid sizes $h_l = 2^{-r} h_{l -1} $ (Fig.~\ref{fig:refine}).  Both the fields $F$ and sources $S$ are defined on different levels of $l$ as $F_{l,r} = F_l$ and $S_{l,r} = S_l$ within a given domain. The coarsest grid corresponds to $l = 1$ and the ratio of the grid size in level $l$ to that in level $1$ can be obtained recursively by $h_l = 2^{-r} h_{l -1}$. Thus, the grid size $h_l$ decreases as $l$ increases. As an example, consider a case where there are 3 levels $\Omega_{1, 1}$, $\Omega_{2, 2}$, $\Omega_{3, 1}$. The cell size for level 3 is $h_3 = h_1 / 8$. The equations to solve at level $l$ can be expressed as:

\begin{equation}\label{eq:FS}
\nabla_{\perp}^2 F_l = S_l,  \nabla_{\perp}^2 F_l - F_l = S_l.
\end{equation}

Here, $S_l$ represents the right-hand side of the equations in (Eq. \ref{eq:poisson}), involving operations associated with the charge and current sources deposited on grid $l$. In the original QuickPIC implementation, the operations $\nabla_{\perp}^2$ are performed in the Fourier space using a Fast Fourier Transform (FFT) library. In the mesh refinement code, the FFT library is still used to manage the $\nabla_{\perp}^2$ operations in equations Eq. \ref{eq:FS} for the coarsest level $l=1$, which encompasses the entire simulation box. For refined meshes with $l > 1$, a finite difference method is applied to handle these operations.

Currently, we maintain uniform grid spacing in both the  $x$ and $y$ directions for each level with $h_l = h_{l,r}$ (and based on our notation, $h_l=2^{-r}h_{l-1}$). We use a first order discretization of the Laplace operator $\nabla_{\perp}^2 F_l$ at grid point $(i,j)$ that takes the form $\nabla_{\perp}^2 F_l = (F_l(i+1,j) + F_l(i-1,j) + F_l(i,j+1) + F_l(i,j-1) - 4 F_l(i,j)) / h_l^2$. We do not define and use intermediate grid points within a cell. For the right hand side of Eq.~\ref{eq:FS}, there will be operations related to $\nabla_{\perp}$, for example divergence or gradient. We use central difference approximations to discretize the derivatives $\nabla_{\perp}$ in the $x$ and $y$ directions for each level $l$. The central difference approximation for the $x$-component of the derivative on grid $(i,j)$ is given by: $\frac{\partial F_l}{\partial x}\bigg|_{(i, j)} \approx \frac{F_l(i+1, j) - F_l(i-1, j)}{2 h_l}$. Similarly, the central difference approximation for the $y$-component of the derivative  on grid $(i,j)$ is given by: $\frac{\partial F_l}{\partial y}\bigg|_{(i, j)} \approx \frac{F_l(i, j+1) - F_l(i, j-1)}{2 h_l}$. For example the divergence of the current in Eq.~\ref{eq:poisson} at level $l$ can be expressed as 

\begin{equation}\label{eq:divj}
\nabla_{\perp} \cdot \vec{J}_{\perp} = \frac{1}{2 h_l} (J_x(i+1,j) - J_x(i-1,j) + J_y(i,j+1) - J_y(i,j-1))
\end{equation}

We show in Fig.~\ref{fig:grid_in} the grid structures near a refinement boundary with a refinement ratio of 2. The deposition of charge density ($\rho$) and current density ($\vec J$) are performed on each refinement level $l>1$ with an extended mesh shown in Fig.~\ref {fig:grid_in} as the dashed-gray grid. For the equations where the gradient is not included in the source term, e.g., ${\nabla_{\perp}^{2} \psi=-\left(\rho-J_{z}\right)}$, the deposited densities $\rho$ and $J$ are only required for the inner grids noted by solid dots. However, when there are terms including $\nabla_{\perp}$ present for the source term of Poisson like equations (e.g., in the equation for $B_z$) , the density $\rho$ and $J$ have to be deposited on the refinement boundary, shown as squares and triangles, which means particles outside the refinement mesh (shown as red dot) contributes to the source terms for the refined mesh. To handle this, we deposit the plasma particle quantities on the fine mesh with the domain extended by one coarse cell (shown as the dashed-gray grid) compared to the fine mesh used to solve the fields.  To improve the efficiency, we choose a strategy where particles are only deposited on the finest extended mesh that encompasses them. Subsequently, the density on lower refinement levels, which overlaps an area with higher refinement levels, is obtained through a coarsening (averaging) process from the meshes of higher levels.

We next describe the structure for solving the discretized equations of Eq.~\ref{eq:FS}. We use the Fast Fourier Transform (FFT) solver from the original QuickPIC to solve for the fields on the coarsest grid, denoted as level $l = 1$, which encompasses the entire simulation box. We then interpolate these fields, denoted as, $F_l$, onto the the boundary of grid level $l+1$ as its Dirichlet boundary condition for $F_{l+1}$. For the refined grids at levels $l > 1$, a geometric multigrid solver (\ref{sec:appendix}) is used. The fields, $F_{l_{max}}$,  for the highest value of $l$ at the location of the particle, 
obtained by solving for the Poisson like equations, 
are used to advance both plasma particles in each 2D step (Eq.~\ref{eq:plasma_push} and \ref{eq:plasma_pz}), as well as beam particles in each 3D step (Eq.~\ref{eq:beam_push}).

\subsubsection{Parallelization schemes in mesh refinement for load balance}

The original QuickPIC utilized a hybrid parallelization approach where MPI is used across nodes and OpenMP is used within a node. This hybrid MPI-OpenMP parallelization together with a pipelining method provides high parallel scalability. Pipelining~\citep{feng2009enhancing} takes advantage of the lack of $\partial/\partial s$ in the field equations and causality. Once a 2D slice at a fixed $s$ has been advanced to a value of $\xi_1$, all the fields required to advance beam particles with $\xi \ll \xi_1$ forward to a new value of $s = s + \Delta s$ are known. At this point a new 2D slice can be initialized at $\xi=0$, and then advanced forward in $\xi$. Thus, a series of ``stages" which correspond to advancing a 2D slice at a different value of $s$ can be processed in parallel. However, when mesh refinement is included in the simulation with high refinement ratios, load balancing issues can arise due to the clustering of a refined mesh to one or two parallel partitions or a small number of stages.

To overcome this load balancing challenge, we employ two strategies. The first strategy focuses on the 2D field solves. In the original QuickPIC implementation, the coarse grid covering the entire simulation domain is divided into multiple partitions along the $y$-direction, with each partition occupying a separate MPI node and maintaining identical physical dimensions. However, in certain scenarios, computational load on the fine meshes may heavily concentrate on only a few MPI nodes, causing long wait time for other MPI nodes. To address this load imbalance issue, a mechanism is implemented: when the number of fine grids surpasses a predetermined threshold, field data and solving processes are redistributed across all processors. After the multigrid cycles are finished, the data is then gathered back to enable the pushing of particles that were not redistributed. The second strategy is aimed at addressing the pipelining algorithm. 
We divide the 3D grids into slabs with nonuniform thicknesses in $\xi$. The thickness of each slab is chosen such that the total number of grids, which includes all cells from the coarsest mesh and all refinement meshes within a slab, is close to the number of grids in every other slab. This technique helps to ensure that the computational load is balanced across the pipelining partitions.


\subsection{Adaptive mesh refinement for evolving beams}

In PWFA simulations, the size of the beam can vary significantly depending on the specific application. For instance, as beams propagate through plasma matching sections, where the plasma density changes adiabatically, the beam size can undergo drastic changes by several orders of magnitude~\citep{PhysRevLett.64.1231, PhysRevAccelBeams.26.121301}.  Even if a beam's spot size is evolving, a static refinement domain can be used where the cell sizes resolve the smallest spot size for the evolving beam and the width of the box is larger than the largest spot size. However, the computational cost can become prohibitively expensive for cases where the beam size changes by more than one order of magnitude. 

To address this issue, we have developed an adaptive mesh refinement scheme that allows the size and resolution of the refined mesh to dynamically adjust according to the beam's changing size. We denote the transverse spot sizes of the beam as $\sigma_x$ and $\sigma_y$, the cell sizes of the finest mesh as $\Delta_x$ and $\Delta_y$, and the dimensions of the box size as $l_x$ and $l_y$. The parameters of the refined mesh (cell size and dimensions) are determined empirically, considering factors such as maintaining result consistency while avoiding anomalies near the boundary. For instance, in scenarios where we simulate ion motion induced by intense beams, we aim to maintain adequate resolution for both the beam size and ion motion while optimizing computational resources. To achieve this, we establish criteria such as $5 \Delta_x \leq \sigma_x \leq 10 \Delta_x $ and $5 \Delta_y \leq \sigma_y \leq 10 \Delta_y$. Additionally, we set the box size to ensure that it surrounds the whole beam while reducing numerical errors. For example, we may use $3 \sigma_x \leq l_x \leq 10 \sigma_x $ and $3 \sigma_y \leq l_y \leq 10 \sigma_y $ as our criteria. These parameters can also be adjusted if a higher resolution is required or if there are abnormalities near the boundary.

\section{Benchmark test of mesh refinement code and performance evaluations}

The mesh refinement code is an effective tool for simulating applications that require ultra-high resolution in specific small regions. However, achieving high refinement ratios can be challenging as simulation errors may be amplified by inaccuracies at the refinement boundaries. Moreover, performing fine resolution everywhere in 3D quasi-static simulations can be computationally expensive and memory-intensive. As a result,  the ``true" results for problems of interest may not be available for comparison. For these reasons, benchmarking the mesh refinement algorithm  must be done  through a variety of methods. 

In this section, we present benchmarks of the mesh refinement code for two bunch PWFA simulations where wakes from electron beam drivers accelerate witness beams of either electrons or positrons. In some cases we conduct simulations using parameters from previous publications. We evaluate the accuracy and efficiency of the mesh refinement code in scenarios requiring ultra-high resolution only in small regions. We consider some cases where there is a high degree of azimuthal symmetry. This allows us to compare mesh refinement results obtained from QuickPIC with those from QPAD, where high resolution is applied across the entire simulation domain. For the positron witness beam case, it is possible to compare the mesh refinement results with a one-step QuickPIC simulation where the finest resolution is used throughout. Additionally, we investigate the mesh refinement algorithm's performance in a scenario featuring a high degree of azimuthal asymmetry in the witness beam's transverse cross-section. This case is also compared with a one-step QuickPIC simulations employing the finest resolution across the entire domain. By carefully assessing the performance of the mesh refinement code for these scenarios, we can better understand its strengths and limitations and identify opportunities for further improvement.

\subsection{A symmetric PWFA simulation with ion motion}\label{sec:sym}
Emittance growth, resulting from the perturbation of focusing forces due to ion motion induced by tightly focused, matched beams, is a topic of importance for plasma based acceleration~\citep{PhysRevLett.95.195002}. This was previously investigated for PWFA in the blowout regime using QuickPIC~\citep{an2017ion}. However, the simulation technique used in that study was not entirely self-consistent. Simulations covering the whole simulation domain were carried out with the finest resolution possible at that time. Based on those results they then isolated the evolution of the witness beam by simulating only a small region of the ion column which contained the witness beam with a very fine resolution, and surrounded it with a ring of immobile negative charge. This permitted an accurate examination for how the the ion motion affected the focusing fields but not the accelerating fields. Within this small region, they were able to use small enough cell sizes to resolve the ion collapse and witness beam spot size. This approach did not take into account the potential effects of the drive beam evolution on the ion motion nor how ion motion affects the accelerating fields.

\begin{figure*}[!ht]
	\begin{center}

\begin{subfigure}{0.45\textwidth}
    \centering 
    \includegraphics[width=1.0\textwidth, height = 0.88\textwidth]{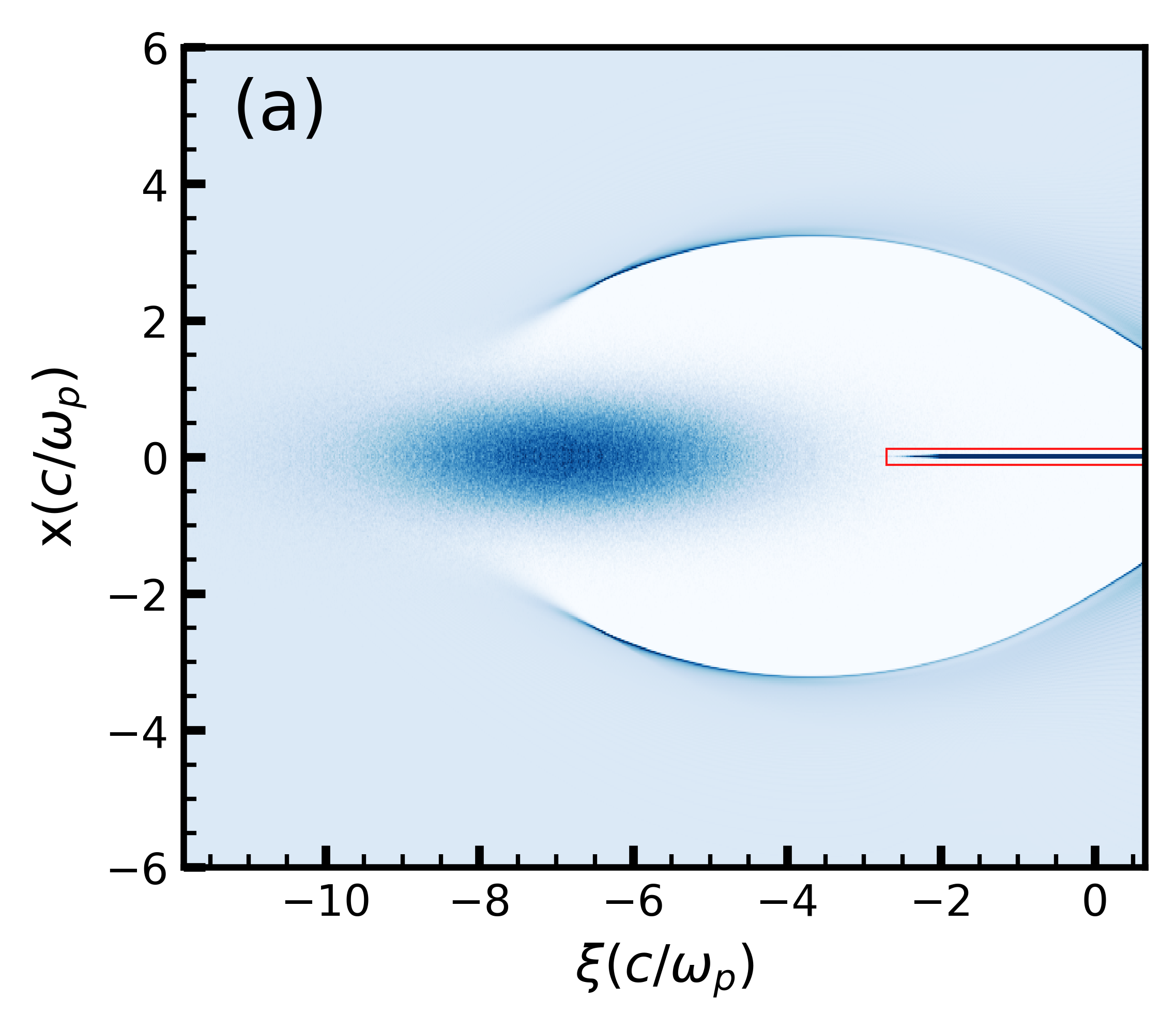}
\end{subfigure} 
 \begin{subfigure}{0.48\textwidth}
 \centering
 \includegraphics[width = 1.0\textwidth, height = 0.8\textwidth]{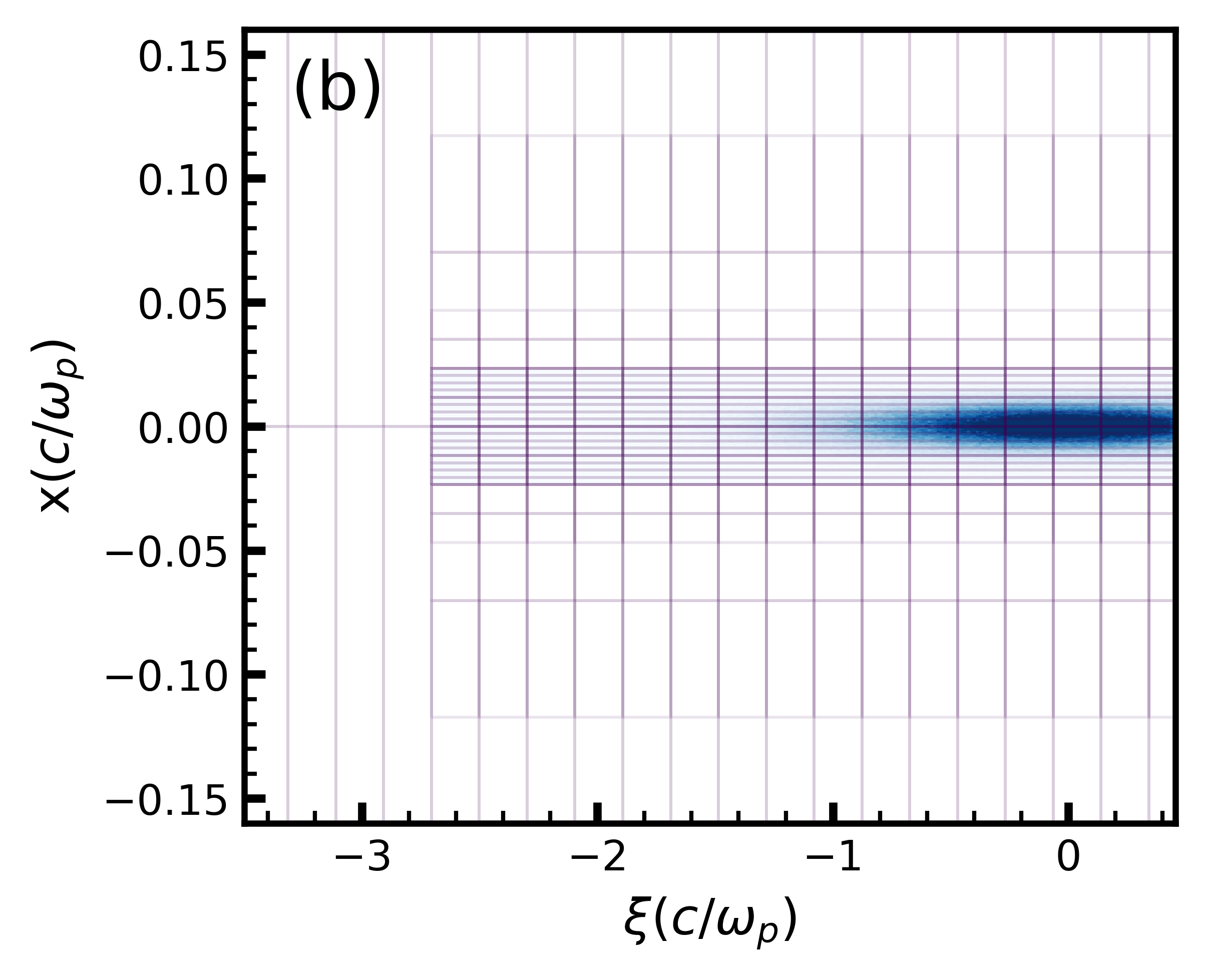}
 \end{subfigure}  
	
		\caption{Setup for a mesh refinement simulation in a PWFA scenario with a symmetric trailing beam. (a) Two-dimensional cut of the three-dimensional charge density including plasma electrons and beams in the $\xi - x$ plane at $y = 0$. The red box surrounding the trailing beam represents the mesh refinement boundary with the highest resolution. (b) The zoomed-in grid setup near the trailing beam. All refinement meshes start at the same $\xi$. As the grid approaches the $x = 0$ axis, the resolution in the $x$ direction progressively increases. In the illustration, for clarity, we present meshes with cell sizes eight times larger in $x$ and sixteen times larger in $\xi$ for each level, compared to the cell sizes employed in the actual simulation.}\label{fig:sym}

	\end{center}
\end{figure*}

We re-examine the symmetric PWFA scenario in~\citep{an2017ion} with fully self-consistent simulations of the entire domain using the mesh refinement method. We simulate the ion motion with the symmetric beam parameters presented in~\citep{an2017ion}, where a matched drive beam and trailing beam (both are bi-Gaussian with definitions of $\sigma_r$ and $\sigma_z$) are loaded in a uniform plasma with density $n_{0}=1.0 \times 10^{17} \mathrm{~cm}^{-3}$ (Fig.~\ref{fig:sym} (a)). The initial energy of the drive beam and trailing beam are both $25~\mathrm{GeV}$.  The plasma skin depth is $k_p^{-1} = 16.8~\mathrm{\mu m}$. The matched drive beam has an rms spot size of $\sigma_{r d}=10.37~\mu \mathrm{m}$ and pulse length of $\sigma_{z d}=30.0~\mu \mathrm{m}$. The trailing beam is loaded $115~\mu \mathrm{m}$ behind the drive beam with a spot size of $\sigma_{r t}=0.1~\mu \mathrm{m}$ and a pulse length of $\sigma_{z t}=10~\mu \mathrm{m}$.

\begin{figure*}[!htbp]
	\begin{center}

\begin{subfigure}{0.48\textwidth}
    \centering 
    \includegraphics[width=1.0\textwidth, height = 0.75\textwidth]{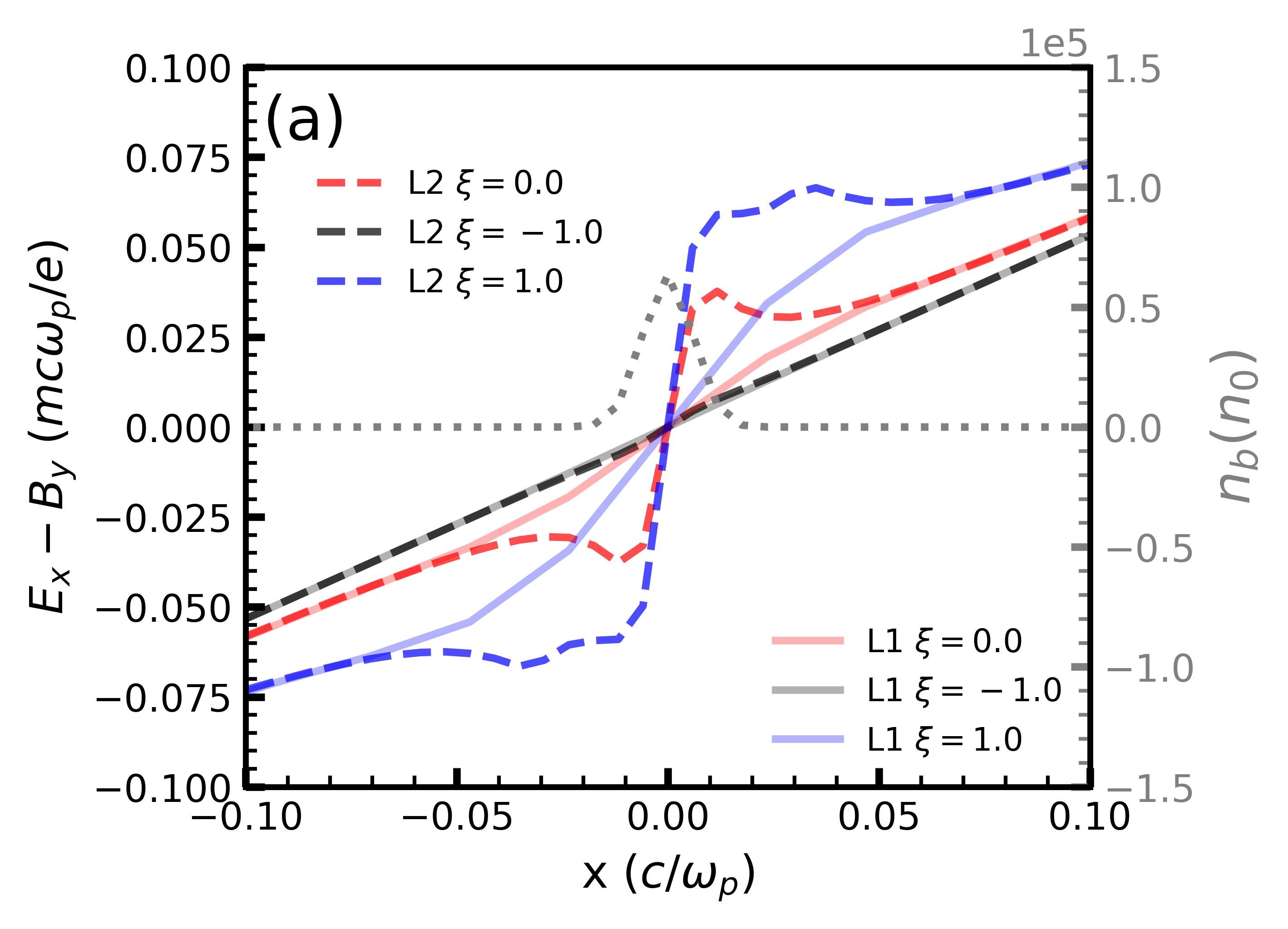}
\end{subfigure} 
 \begin{subfigure}{0.48\textwidth}
 \centering
 \includegraphics[width = 1.0\textwidth, height = 0.75\textwidth]{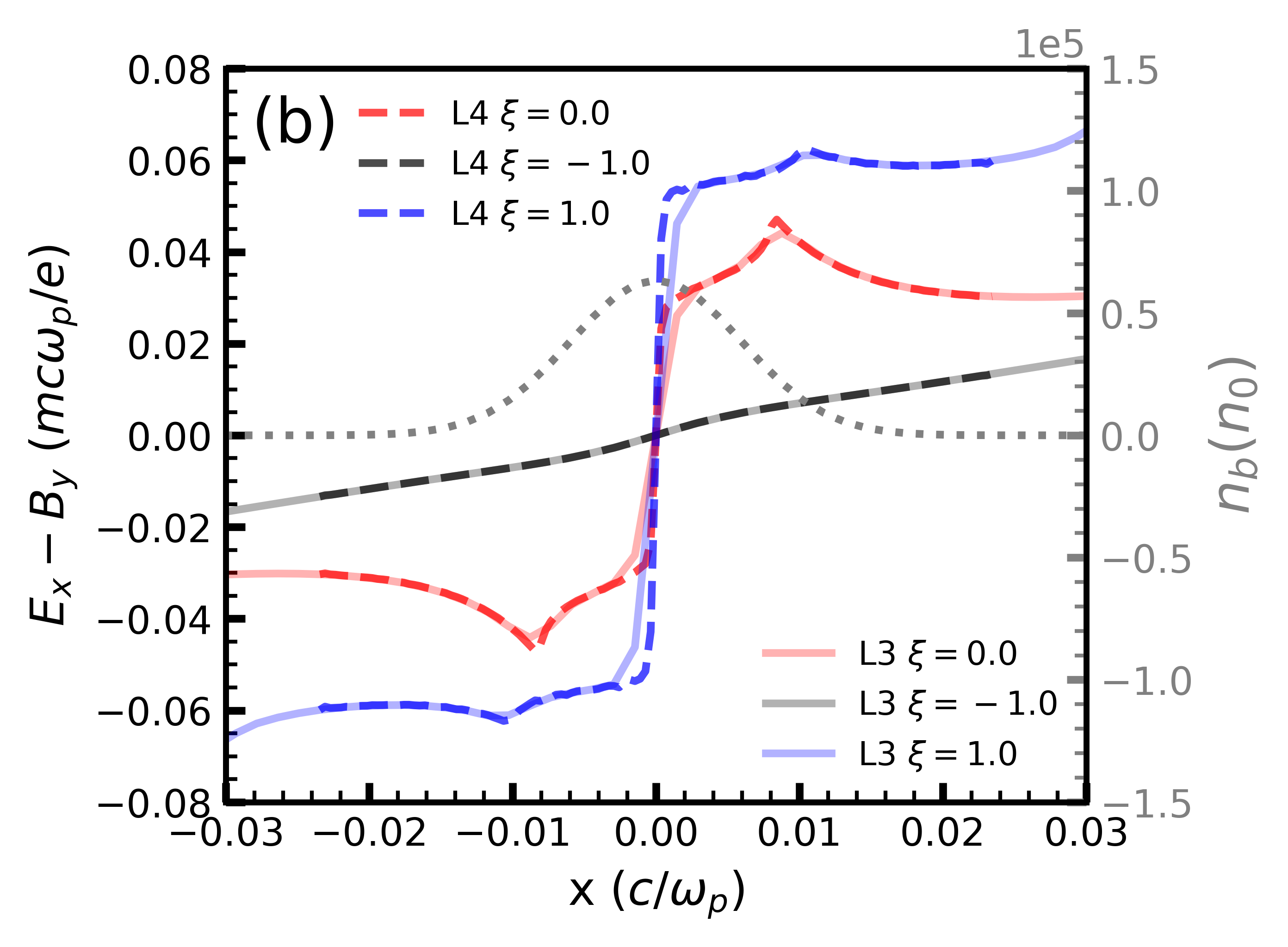}
 \end{subfigure}  

  \begin{subfigure}{0.48\textwidth}
 \centering
 \includegraphics[width = 1.0\textwidth, height = 0.75\textwidth]{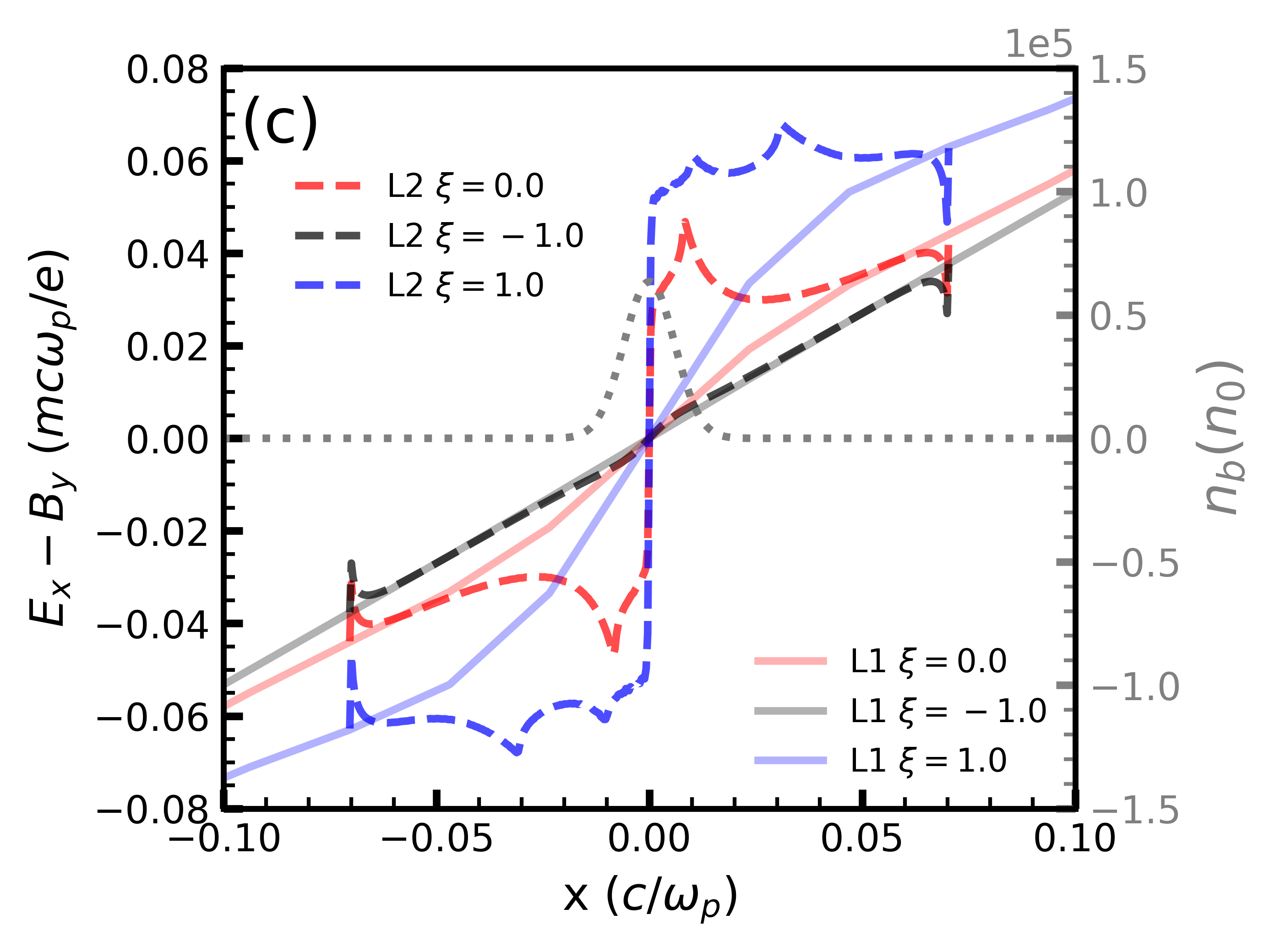}
 \end{subfigure}  
\begin{subfigure}{0.45\textwidth}
 \centering
 \includegraphics[width = 1.0\textwidth, height = 0.85\textwidth]{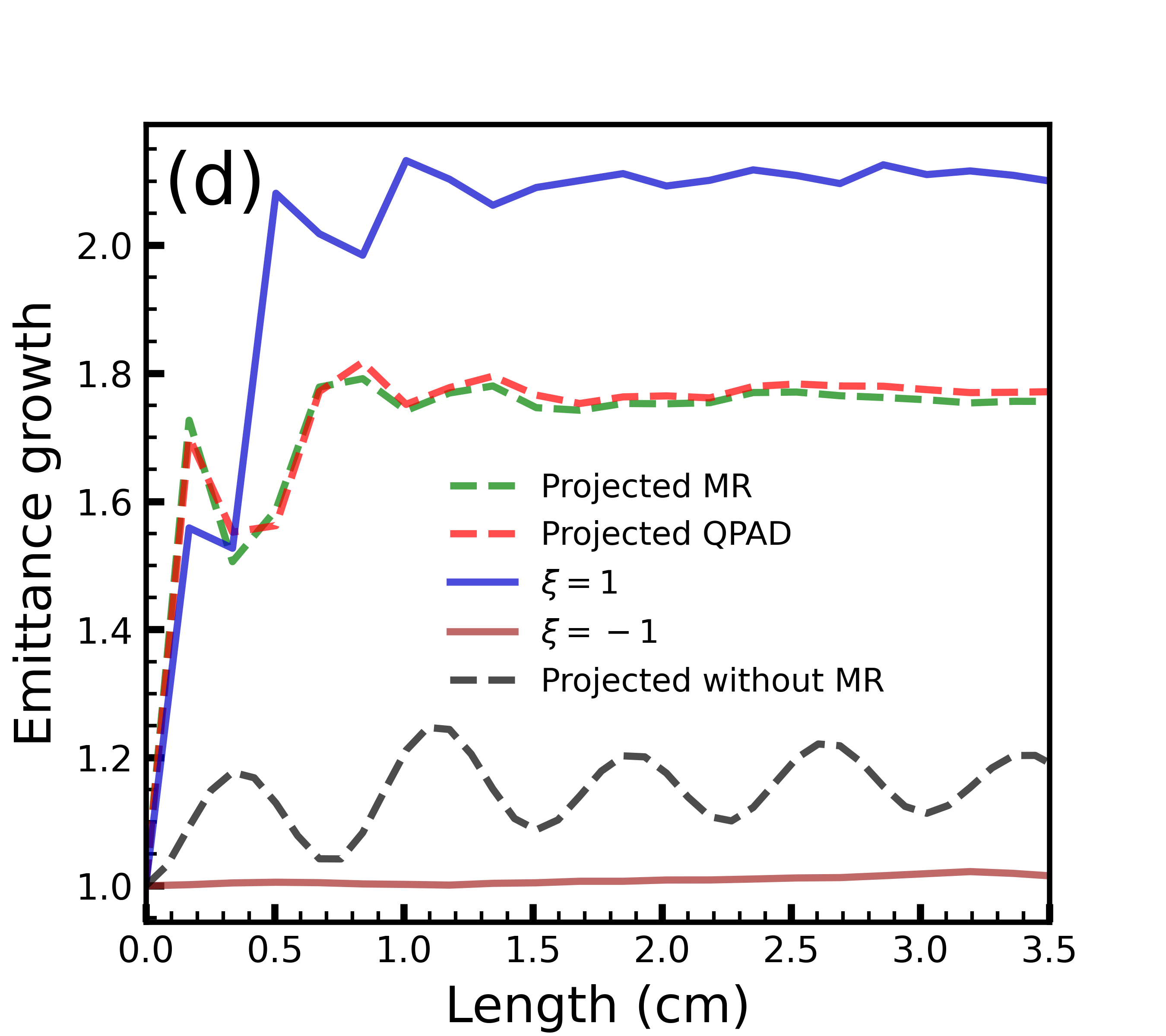}
 \end{subfigure}  
		
		\caption{Simulation results using the setup depicted in Fig.~\ref{fig:sym}. (a) Lineouts of the focusing field $E_x - B_y$  at $y = 0$ solved by the 4-level mesh refinement at different $\xi$ values. We present the focusing fields for the coarsest level (L1) $\Omega_{1,1}$ and refinement level 2 (L2) $\Omega_{2,2}$. The trailing beam density is represented by the dotted-gray line. The focusing fields of level 1 (L1) at $\xi = -1.0, 0.0, 1.0$ are illustrated by the gray, red, and blue lines, respectively. The focusing fields of level 2 (L2) at $\xi = -1.0$ (dashed black), $\xi = 0.0$ (dashed red), and $\xi = 1.0$ (dashed blue) are also depicted. (b) Lineouts of the focusing field $E_x - B_y$  at $y = 0$ for refinement levels 3 (L3) $\Omega_{3,2}$ and 4 (L4) $\Omega_{4,2}$. The dotted-gray line represents the trailing beam density. The dashed lines represent focusing fields of level 4 (L4) at $\xi = -1.0~\text{(dashed black)}, 0.0~\text{(dashed red)}, 1.0~\text{(dashed blue)}$. The gray line, red line and blue line represent the focusing field of level 3 (L3) at $\xi = -1.0, 0.0, 1.0$ respectively. (c) Lineouts of the focusing field $E_x - B_y$  at $y = 0$ solved by mesh refinement with only 2 levels at different $\xi$ values. Here we show focusing fields of the coarsest level (L1) $\Omega_{1, 1}$ and refinement level 2 (L2) $\Omega_{2,6}$. It's noteworthy that level 2 has the same resolution as level 4 in the previous 4-level refinement simulation. The dotted-gray line represents the trailing beam density. The gray line, red line and blue line represent the focusing fields of level 1 (L1) at $\xi = -1.0, 0.0, 1.0$ respectively. The dashed lines represent focusing fields of level 2 (L2) at $\xi = -1.0~\text{(dashed-black)}, 0.0~\text{(dashed-red)}, 1.0~\text{(dashed-blue)}$. With only one level of high resolution refinement, the field near the refinement boundary show abnormalities. (d) The emittance evolution of the trailing beam using mesh refinement, both projected (dashed-green) and in slices at $\xi = -1$ (brown) and $\xi = 1.0$ (blue), is shown. The projected emittance growth of the trailing beam, simulated by QPAD, is presented as the dashed-red line, and QuickPIC coarse resolution without mesh refinement is depicted by the dashed-black line. }\label{fig:sym2}
	\end{center}
\end{figure*}

For the coarsest grid, we use a simulation box with $1024 \times 1024 \times 1024$ cells and a $ \pm 12.0 \times \pm 12.0 \times 13.0~k_{p}^{-1}$ physical dimension in $x,y$, and $\xi$, respectively. The resolution of the coarse grid is $0.0234 \times 0.0234 \times 0.0127~k_{p}^{-1}$, corresponding to $393.86 \times 393.86 \times 213.34 \mathrm{~nm}$. This resolution is insufficient to resolve the spot size of the trailing beam. The finest resolution of the refined grid is $6.15 \times 6.15 \times 213.34 \mathrm{~nm}$, where the transverse grid size is $64$ times smaller than the coarsest grid. This finer resolution matches the resolution used in the isolated small domain described in~\citep{an2017ion} .

We use four refinement levels denoted by $\Omega_{1,1}, \Omega_{2,2}, \Omega_{3,2}, \Omega_{4,2}$ to gradually change the resolutions to $h_{4,2} = h_{1,1} / 64$ (Fig.~\ref{fig:sym} (b)). We define the trailing beam center as $(0.0, 0.0, 0.0)$ in Cartesian coordinates. The domain of the grid with the finest resolution $\Omega_{4,2}$ extends to $\pm 0.023~k_{p}^{-1}$ in both $x$ and $y$ direction, and starts from $-2.7~k_{p}^{-1}$ and ends at $1.15~k_{p}^{-1}$ in the $\xi$ direction. The other refinement levels are initialized to have the same dimension in $\xi$. The dimensions of the second refinement level $\Omega_{2,2}$ are $\pm 0.12~k_{p}^{-1}$ in the $x$ and $y$ directions, respectively. The third refinement level $\Omega_{3,2}$ has dimensions of $\pm 0.047~k_{p}^{-1}$ in the $x$ and $y$ directions.

Figs.~\ref{fig:sym2} (a) and (b) present the focusing field $E_x - B_y$ in the $x$ direction for various slices in $\xi$ obtained using mesh refinement with multiple refinement levels $\Omega_{1,1}, \Omega_{2,2}, \Omega_{3,2}, \Omega_{4,2}$. The fields correspond to when the drive beam has propagated one $s$ time step, $10.0~k_p^{-1}$, into the plasma. As can be seen, from Fig.~\ref{fig:sym2} (b), the values for $E_x - B_y$  and its gradient both smoothly match between levels 3 and 4. To illustrate the benefits of employing multiple refinement levels with progressively changing resolutions, we also present a one step simulation in Fig.~\ref{fig:sym2} (c) featuring only two levels $\Omega_{1,1}, \Omega_{2,6}$. The resolution of the fine mesh is identical to that of L4 shown in Fig.~\ref{fig:sym2} (b). In this case, spurious fields near the refinement boundary are observed (Fig.~\ref{fig:sym2} (c) near $x=\pm 0.07~c/\omega_p$). Recall that in QuickPIC, the focusing field of beam particles is indirectly solved by the gradient of $\psi$: $\vec{E}_{\perp}+\hat z \times \vec {B}_{\perp}=-\nabla_{\perp} \psi$, with $\psi$ solved by $\nabla_{\perp}^{2} \psi=-\left(\rho-J_{z}\right)$. As described above, we only match the Dirichlet boundary for each level of refinement when solving $\psi$; thus the gradient is not matched. We note that we have experimented with iterating back and forth between  coarse and fine meshes such that both Dirichlet and Neumann conditions are satisfied. This does lead to smaller spurious signals near the boundary, however, this requires a more complicated source deposit. We find that the use of Dirichlet boundary conditions works well particularly if the boundaries are chosen far enough away from high order fields or sources. With only two refinement levels, bilinear interpolation at the boundary can introduce significant errors, resulting in the abnormal fields near the refinement boundary as shown in Fig.~\ref{fig:sym2} (c). Therefore, utilizing multiple refinement levels provides more accurate solutions near the boundary and reduces errors caused by discrepancies in the field gradients. We note that for the case with 4 levels the results within the L4 domain in Fig.~\ref{fig:sym2} (b)  look similar to those for the L2 domain in Fig.~\ref{fig:sym2} (c) except at the boundary. Hence, by setting the refinement boundary far away from the trailing beam, the wakefield near the beam for only two levels can closely approximate what we observe with the four levels of refinement. 

Additionally, in Fig.~\ref{fig:sym2} (d) we present the emittance growth over long propagation distances obtained using four levels of mesh refinement $\Omega_{1,1}, \Omega_{2,2}, \Omega_{3,2}, \Omega_{4,2}$, and the same setup as shown in Figs.~\ref{fig:sym2} (a)(b). Using QPAD, we can simulate the entire domain using the finest resolution. All normalized beam and plasma parameters are kept the same, and the agreement can be seen in Fig.~\ref{fig:sym2} (d) where the emittance growth from the QPAD simulation is shown. Through self-consistent mesh refinement simulations, we obtained emittance evolution that is also consistent with the results from~\citep{an2017ion}, that only modeled a small region of the ion channel. It is noteworthy that when employing only the coarse mesh $\Omega_{1,1}$ without mesh refinement, the emittance growth from ion motion cannot be fully resolved, as indicated by the dashed-black line in Fig.~\ref{fig:sym2} (d).

\subsection{PWFA simulation with ion motion for an asymmetric trailing beam }\label{sec:asy}
As mentioned above, for simulations characterized by high degrees of azimuthal symmetry, the quasi-3D QS code  QPAD, can be used to greatly reduce the computational requirements by truncating the number of azimuthal modes to a low number. Additionally, fine resolution can be used throughout in QPAD. However, in certain PWFA applications, such as those involving linear collider parameters, the disparity in spot sizes between different transverse directions can be significant. In such scenarios, a quasi-3D QS code like QPAD may necessitate more than ten azimuthal modes, resulting in slower computation speeds and increased memory requirements. 

With mesh refinement, the computational time and memory costs for a 3D quasi-static simulation need not increase significantly if the refined grids remain confined to a smaller region. We next benchmark the mesh refinement algorithm in a case where the trailing beam exhibits large asymmetry and significant ion motion is triggered. We compare the emittance growth from the mesh refinement simulation, where the entire wakefield is modeled, with the results from the reduced-domain simulation as presented in~\citep{an2017ion}. Additionally, we compare the focusing fields at a single 3D time step with those obtained from a QuickPIC simulation where the finest grid size is applied throughout the entire simulation domain. We note that this is feasible because in this case the finest resolution is coarser than that required for the symmetric case described earlier as the ion collapse is not as severe or narrow as was the case for the symmetric case. 

\begin{figure*}[!ht]
	\begin{center}

\begin{subfigure}{0.43\textwidth}
    \centering 
    \includegraphics[width=1.0\textwidth, height = 0.8\textwidth]{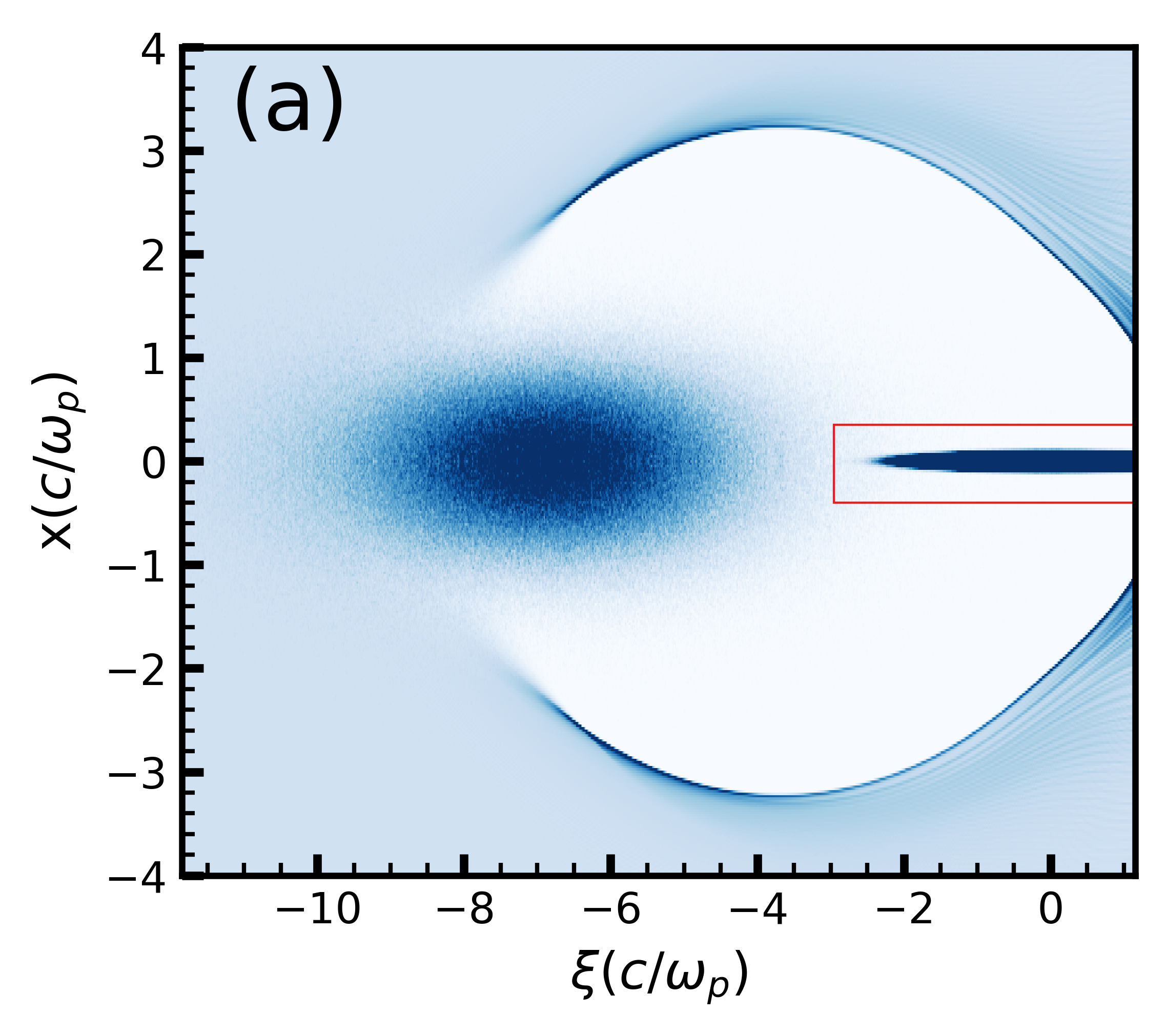}
\end{subfigure} 
 \begin{subfigure}{0.43\textwidth}
 \centering
 \includegraphics[width = 1.0\textwidth, height = 0.8\textwidth]{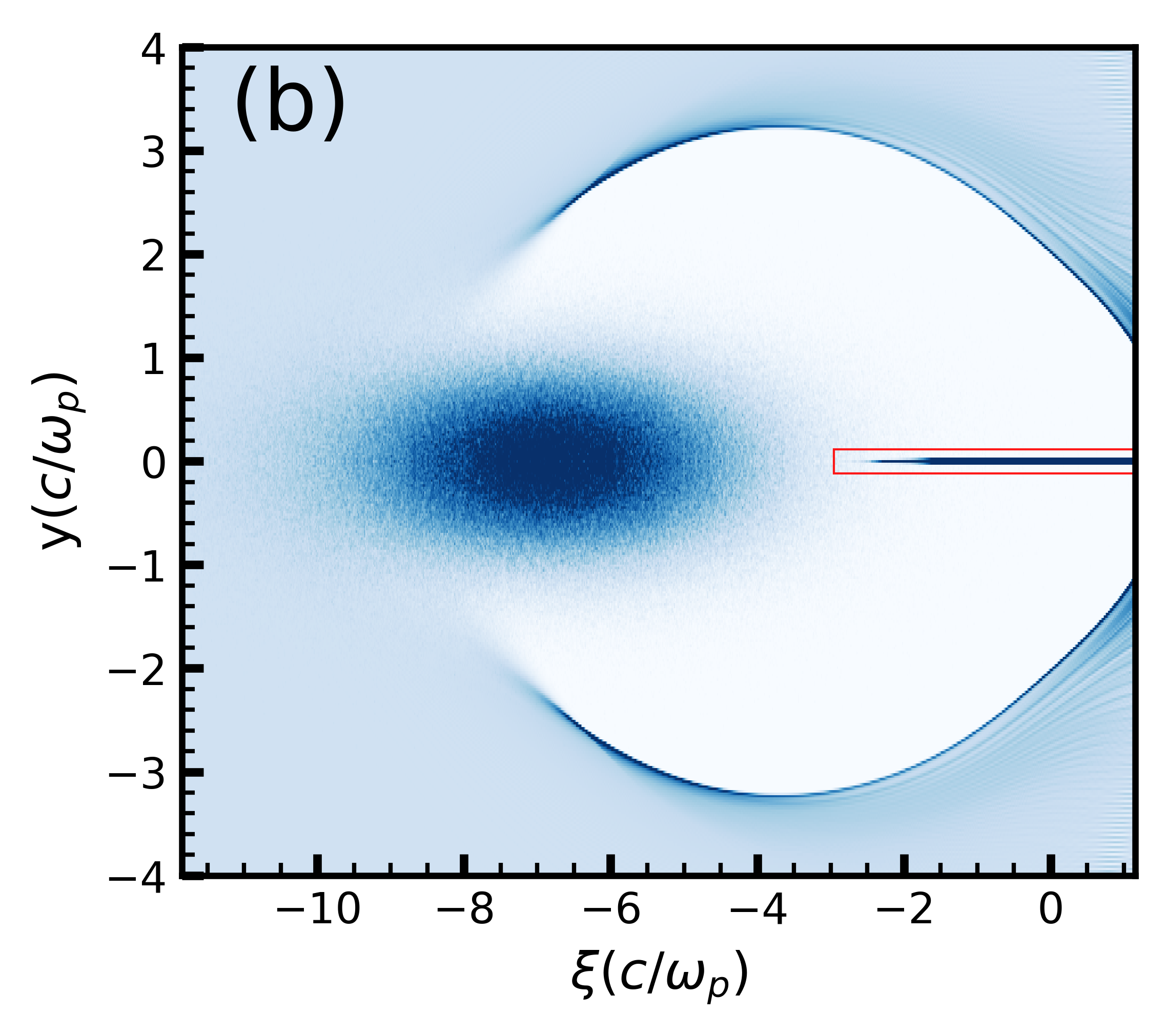}
 \end{subfigure}  
		
		\caption{PWFA simulation with mesh refinement for an asymmetric trailing beam. (a) Two-dimensional $\xi - x$ slices of the three-dimensional data at $y = 0$ for the charge density of plasma electrons excited by a bi-Gaussian drive beam and an asymmetric trailing beam. The red box surrounding the trailing beam represents the mesh refinement boundary with the highest resolution. (b) Two-dimensional $\xi - y$ slices of the three-dimensional data at $x = 0$ for the charge density of plasma electrons excited by a bi-Gaussian drive beam and an asymmetric trailing beam. The red box surrounding the trailing beam represents the mesh refinement boundary with the highest resolution. }
            \label{fig:asy}
	\end{center}
\end{figure*}

\begin{figure*}[!ht]
	\begin{center}

\begin{subfigure}{0.42\textwidth}
    \centering 
    \includegraphics[width=1.0\textwidth, height = 0.8\textwidth]{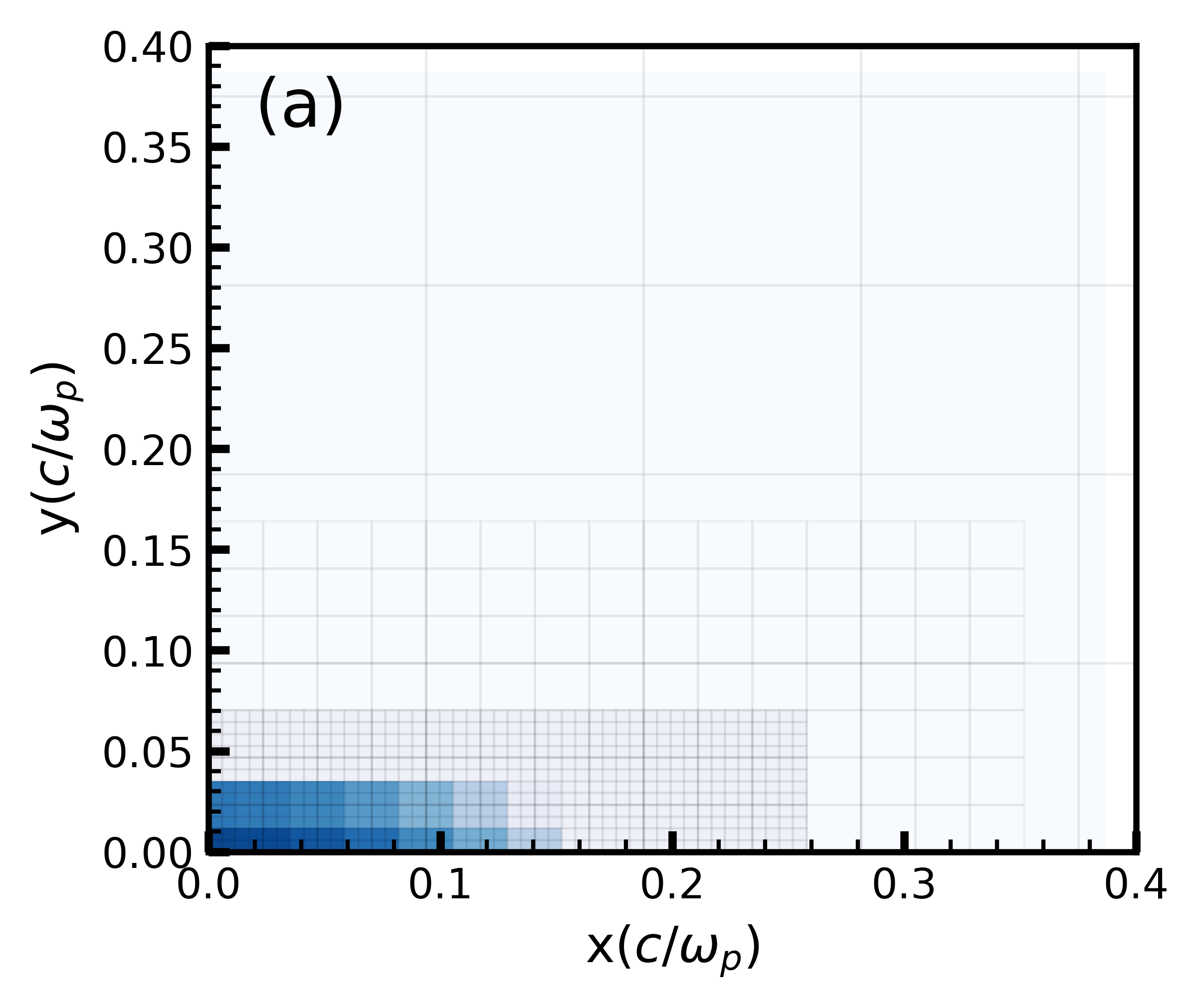}
\end{subfigure} 
 \begin{subfigure}{0.44\textwidth}
 \centering
 \includegraphics[width = 1.0\textwidth, height = 0.78\textwidth]{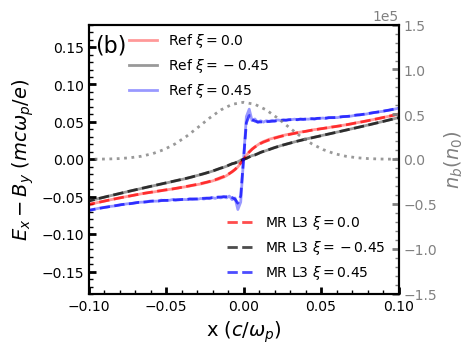}
 \end{subfigure}  

  \begin{subfigure}{0.44\textwidth}
 \centering
 \includegraphics[width = 1.0\textwidth, height = 0.78\textwidth]{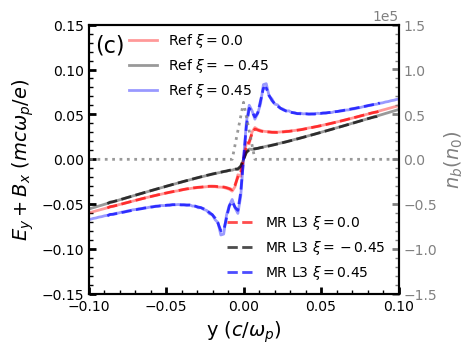}
 \end{subfigure}  
\begin{subfigure}{0.43\textwidth}
 \centering
 \includegraphics[width = 1.1\textwidth, height = 0.8\textwidth]{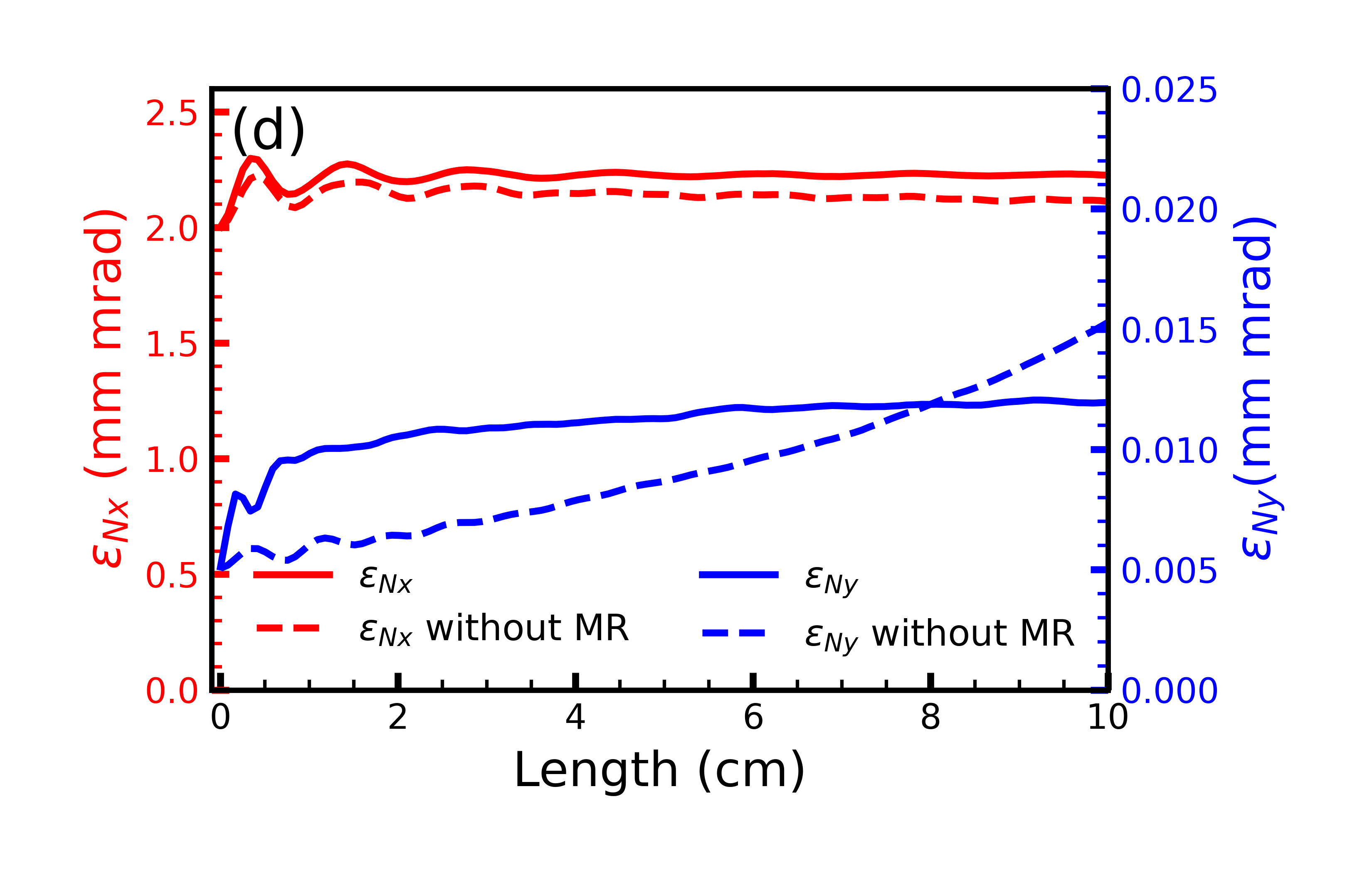}
 \end{subfigure}  
		
		\caption{This figure illustrates the refined meshes and presents simulation results using the setup depicted in Fig.~\ref{fig:asy}. (a) A zoomed-in slice in the $x-y$ plane at $\xi = 0$ depicting the trailing beam density of level 1. A quarter of the beam is shown, with symmetry along both the $x$ and $y$ axes. We also show the refined grid with levels of progressively varying resolutions that covers the trailing beam. To distinctly illustrate the nesting of levels in the figure, we present meshes with cell sizes four times larger for each level compared to the sizes employed in the simulation. (b) Lineouts of the focusing field $E_x - B_y$ along $x$ at $y = 0$ for various slices of the trailing beam from a  mesh refinement simulation. The dotted-gray line represents the beam density of the trailing beam. The focusing field at slices $\xi =$ -0.45 , 0.0 ,  0.45 is shown as dashed-black, dashed-red and dashed-blue line for mesh refinement solution at level 3, which agrees with the fine-resolution everywhere QuickPIC solution at $\xi =$ -0.45 (gray) , 0.0 (red),  0.45 (blue). (c) Lineouts of the focusing field $E_y + B_x$ along $y$ at $x = 0$ for various slices of the trailing beam from a mesh refinement simulation. The dotted-gray line represents the beam density of the trailing beam. The focusing field at slices $\xi =$ -0.45 , 0.0 ,  0.45 is shown as dashed-black, dashed-red and dashed-blue line for mesh refinement solution at level 3, which agrees with the fine-resolution everywhere QuickPIC solution at $\xi =$ -0.45 (gray) , 0.0 (red),  0.45 (blue).  (d) Normalized emittance growth in the $x$ plane, denoted as $\epsilon_{Nx}$ (red line), and in the $y$ plane, denoted as $\epsilon_{Ny}$ (blue line). Additionally, the normalized emittance growth simulated with coarse resolution without mesh refinement is shown as dashed lines (red in the $x$ plane, blue in the $y$ plane). }
            \label{fig:asy2}
	\end{center}
\end{figure*}

We perform a PWFA simulation with the same drive beam parameters and plasma density as the symmetric PWFA simulation in Sec.~\ref{sec:sym}, but with an asymmetric trailing beam. The same parameters were also investigated in~\citep{an2017ion}. The center of the trailing beam is loaded $115~\mathrm{\mu m}$ behind the center of the drive beam and initialized with emittances $\epsilon_{N x}=2.0 \mathrm{~\mu m}, \epsilon_{N y}=0.005 \mathrm{~ \mu m}$ in the transverse plane. The initial spot sizes are $\sigma_{x}=463.9 \mathrm{~nm}$ and $\sigma_{y}=23.2 \mathrm{~nm}$ (Figs.~\ref{fig:asy} (a) and (b)) which are the matched spot sizes for the unperturbed focusing field. The same simulation box with identical cells and physical dimensions as described in Sec.~\ref{sec:sym} is used for the coarsest grid. In this simulation we use three refinement levels $\Omega_{1,1}, \Omega_{2,2}, \Omega_{3,2}$ where the grid is shown in Fig.~\ref{fig:asy2} (a), and the refinement domains are chosen so that increasing resolution or the physical dimension does not impact the results within the desired accuracy. The finest grid resolution is $16$ times smaller than the coarsest grid resolution and has a rectangular shape where the grid boundaries are at $lx, ly = \pm 0.26 , \pm0.07 ~k_p^{-1}$. The resolution of the finest mesh is $0.0015~k_p^{-1}$, which resolves the smaller transverse spot size of the trailing beam.

Figs.~\ref{fig:asy2} (b) and (c) present the focusing field $E_x - B_y$ and $E_y + B_x$ obtained using mesh refinement on $\Omega_{3,2}$ for various slices in $\xi$. The fields correspond to when the drive beam has propagated one $s$ time step $10.0~k_p^{-1}$ into the plasma. The results agree with the full 3D QuickPIC simulation that used the finest resolution $0.0015~k_p^{-1}$ uniformly throughout the simulation window. The emittance growth of the trailing beam in the x and y directions is shown in Fig.~\ref{fig:asy2} (d), where the larger emittance in $x$, $\epsilon_{N x}$, grows by only $11 \%$, and $\epsilon_{N y}$ grows by $155 \%$. The emittance growth we obtained is slightly higher than but still close to that found in the high resolution Ref.~\citep{an2017ion}, validating the effectiveness of mesh refinement code for highly asymmetric problems. We note that in both the symmetric and asymmetric witness beam simulations, we did not incorporate quasi-adiabatic matching  sections~\citep{PhysRevAccelBeams.26.121301}, which can mitigate emittance growth effectively.

\subsection{Simulation of plasma-based positron acceleration}\label{sec:positron}

Simulating plasma-based positron acceleration with mesh refinement introduces additional challenges. Several PBA schemes have been proposed to accelerate positron beams in plasma~\citep{PhysRevAccelBeams.22.081301, PhysRevLett.127.104801,PhysRevLett.127.174801,PhysRevAccelBeams.25.091303, PhysRevLett.112.215001}. In many of these regimes, the focusing field is generated by a narrow electron filament that forms along the axis. Setting up a refinement mesh around the positron beam can have a large cluster of particles crossing the refinement boundary. Previous studies have shown that particles passing the refinement boundary can suffer from spurious self-forces~\citep{vay2002mesh}. In this study, we do not explicitly examine the self-force issue that arises as particles cross the boundary; rather we mitigate spurious boundary effects by carefully establishing a hierarchy of resolution layers and positioning the refinement boundary away from clusters of a large number of particles. Subsequently, we compare the fields solved by mesh refinement with those obtained using the QuickPIC where a fine mesh is used for the entire simulation domain.

\begin{figure*}[!ht]
	\begin{center}

\begin{subfigure}{0.45\textwidth}
    \centering 
    \includegraphics[width=1.0\textwidth, height = 0.8\textwidth]{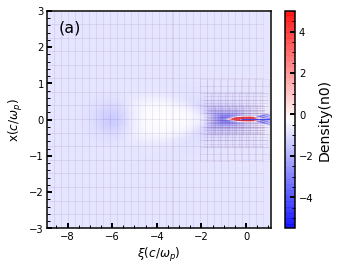}
\end{subfigure} 
 \begin{subfigure}{0.45\textwidth}
 \centering
 \includegraphics[width = 1.0\textwidth, height = 0.8\textwidth]{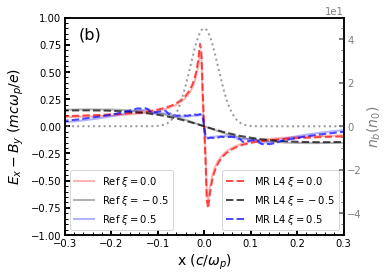}
 \end{subfigure}  

   \begin{subfigure}{0.44\textwidth}
 \centering
 \includegraphics[width = 1.0\textwidth, height = 0.78\textwidth]{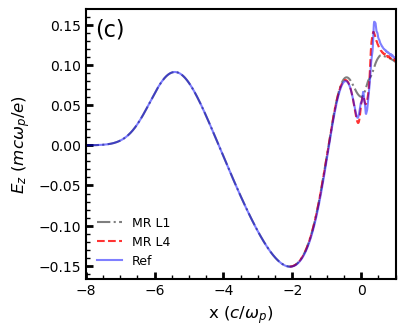}
 \end{subfigure} 
		
    \caption{Mesh refinement simulation results of positron acceleration using a two-beam setup, the one-step results are compared with a fine-resolution everywhere QuickPIC simulation. (a) A cross-section of the three-dimensional charge density data for plasma electrons excited by a bi-Gaussian drive electron beam centered at $\xi = -5.9$ and a trailing positron beam loaded at $\xi = 0.0$. The charge density of both the drive beam and the following beam is also illustrated. (b) The focusing field for various slices of the trailing beam solved by mesh refinement. The dotted-gray line represents the beam density of the trailing beam. The focusing field at slices $\xi =$ -0.5, 0.0, 0.5 is shown as dashed-black, dashed-red and dashed-blue lines for the mesh refinement solution at level 4, which agrees with the fine-resolution everywhere QuickPIC solution at $\xi =$ -0.5 (gray), 0.0 (red),  0.5 (blue). (c) On-axis electric field $E_z (\xi)$ of the coarsest level (L1) $\Omega_{1, 1}$ (dash-dotted gray) and finest level (L4) $\Omega_{4, 1}$ (dashed red) simulated by mesh refinement, the results agree with the fine-resolution everywhere QuickPIC solution (blue).}
            \label{fig:positron}
	\end{center}
\end{figure*}

We present results on the use of mesh refinement in simulating positron acceleration using a two-beam setup consisting of an electron drive beam and a positron trailing beam. In Fig.~\ref{fig:positron} (a) we present a slice in the $\xi-x$ plane for the the total density (blue is net negative and red is net positive) for a two-beam simulation where the drive beam has a bi-Gaussian profile centered at $k_p \xi = -5.9$ with $k_p \sigma_z = 0.6$, $k_p \sigma_r = 0.3$ and peak density $n / n_0 = 1.0$. The trailing beam has a bi-Gaussian profile and is loaded at $k_p \xi = 0.0$ with $k_p \sigma_z = 0.4$, $k_p \sigma_r = 0.03$ and peak density $n / n_0 = 50.0$. The simulation window has a size of $12.0 \times 12.0 \times 10.0~k_{p}^{-1}$ in $x, y, z$, and the whole simulation domain is discretized into $512 \times 512 \times 1024$ coarse grids. We use four levels of refinement $\Omega_{1,1}, \Omega_{2,1}, \Omega_{3,1} \text{ and } \Omega_{4,1}$. The refinement meshes are shown in Fig.~\ref{fig:positron} (a) and the resolution in $x$ and $y$ is progressively increased to $0.003~k_p^{-1}$ that is $8$ times smaller than the coarsest resolution. In Fig.~\ref{fig:positron} (b) we show lineouts along $x$ of the the focusing field, $E_x - B_y$, for axial slices at $\xi = -0.5 \text{ (dashed-black)}, 0.0 \text{ (dashed-red)}, \text{ and } 0.5 \text{ (dashed-blue)}$. A lineout of the positron beam density profile (dotted-gray) at $\xi = 0.0$ is shown for reference. Fig.~\ref{fig:positron} (c) shows the on-axis electric field $E_z (\xi)$ of the coarsest level (L1) $\Omega_{1, 1}$ (dash-dotted gray) and finest level (L4) $\Omega_{4, 1}$ (dashed red) simulated by mesh refinement. Without mesh refinement, the coarse resolution fails to resolve the narrow spike of electrons drawn to the axis, with a width of less than $0.005$. With four levels of refinement, both the focusing field and accelerating field agree with those obtained from a 3D QuickPIC simulation where fine resolution is used throughout, which is referred to as the reference case.

\subsection{Adaptive mesh refinement for an evolving beam}
\begin{figure}[!h]
	\centering
	\includegraphics[width=1.0\linewidth]{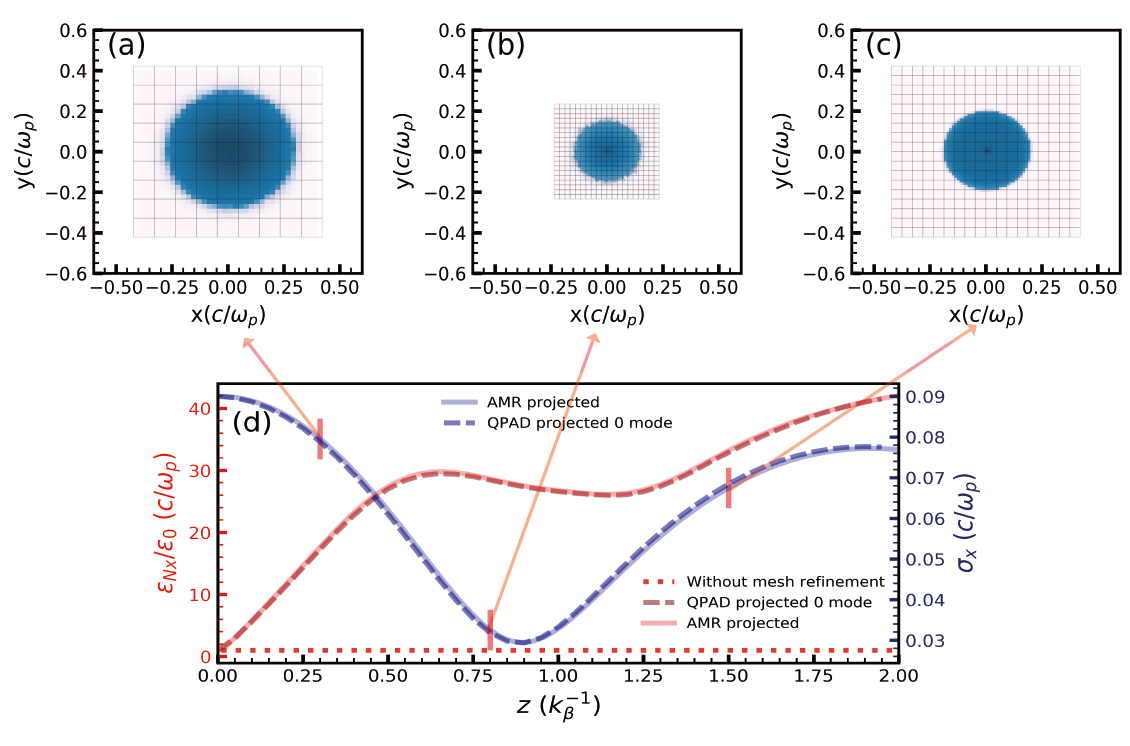}
	\caption{The plot shows the evolution of emittance growth ($\epsilon_{Nx}/\epsilon_0$, indicated by the red line, $\epsilon_0$ is the initial normalized emittance) and bunch size ($\sigma_x$, represented by the blue line) of an unmatched trailing beam. The emittance growth (red line) and evolving spot size  (blue line) simulated by the mesh refinement are compared with those obtained from QPAD (dashed-red line for emittance growth and dashed-blue line for spot size). The projected beam density on the transverse plane at $\xi = 0.3~k_{\beta}^{-1} \text{(a)}, 0.8~k_{\beta}^{-1}\text{(b)}, 1.5~k_{\beta}^{-1}\text{(c)}$ is also presented (we utilize a logarithmic scale for the particle density to visualize not only the high-density beam regions but also the low-density beam boundary). As the beam size changes, the refinement box size and resolution dynamically evolve to encompass the beam while simultaneously resolving the ion motion. The evolution of grids at the highest refinement level is shown on top of the beam density, highlighting the changes in mesh boundary and resolution corresponding to the beam size. }\label{fig:amr}
\end{figure}

To benchmark the adaptive mesh refinement module for evolving beams, we use the same parameter set described in Section \ref{sec:sym} that produced Figs.~\ref{fig:sym} and \ref{fig:sym2}. The position and charge of the trailing beam remain unchanged. However, instead of using a matched trailing beam with a spot size of $0.1~\mathrm{\mu m} = 0.006~k_p^{-1}$, we use an initially unmatched monoenergetic trailing beam with a relativistic factor of $\gamma = 5000$ and a spot size of $\sigma_r = 0.09~k_p^{-1}$, along with a momentum divergence of $\sigma_p = 0.1~m_e c$, where $m_e$ is the electron mass. In this setup, we set the ion mass to be $200~m_e$ instead of the proton mass $1836~m_e$, relaxing the resolution requirement to resolve the ion motion for benchmarking. The spot size of the unmatched beam will evolve dramatically during the simulation (blue curve in Fig.~\ref{fig:amr} (d)). It reaches its minimum value at approximately $0.03~k_p^{-1}$. The refinement box size and resolution will evolve with the beam size to include the beam and at the same time resolve the ion motion. In the simulation, the smallest grid size of the refinement mesh is $0.003~k_p^{-1}$. The evolution of the region with the finest grid is shown in Figs.~\ref{fig:amr} (a) (b) and (c). This case has been tested on a local desktop, where a full resolution QuickPIC simulation would not be possible and instead it was run on a supercomputer. The evolution of emittance growth and spot size is shown in Fig.~\ref{fig:amr} (d). The results agree with a 0-mode QPAD simulation where a fine mesh with a resolution of $0.003~k_p^{-1}$ is used throughout the entire domain and for all time steps. 

\subsection{Performance evaluation}
The mesh refinement code will accelerate the time to solution for the 3D simulation in QuickPIC due to the decrease in computational load from a reduction in cells in the transverse ($x$ and $y$) and longitudinal ($\xi$) directions. However, quantifying the speed up is complicated as the algorithm includes field solves, particle pushes, and iterations. Furthermore, the number of particles per cell does not need to be kept constant as one decreases the cell sizes. The speedup of mesh refinement code highly depends on the scale difference of the physics problem, and has the best performance where the simulation only requires high resolution at a small region and can use a coarser resolution elsewhere, and where the refinement ratio is large. However, for such a problem, the nonuniform grid in mesh refinement will influence the load balance and limit the parallelization scaling in the transverse direction. As mentioned before, we distribute the refined meshes to other MPI ranks and use pipelining to boost the performance of these problems.  

To assess the potential speedup achievable with the mesh refinement approach, we examine several scenarios where uniform-grid QuickPIC simulations can be conducted across the entire simulation domain. The estimated speedup is based on simulations tested on NERSC Perlmutter nodes. In Section~\ref{sec:asy}, we conducted simulations involving an asymmetric electron witness beam inducing ion motion. Employing mesh refinement with a maximum refinement ratio of 16 and $1024 \times 1024 \times 1024$ cells for the coarsest resolution, we achieved speedup of around 350 compared to a fully resolved QuickPIC simulation for a single $s$ step simulation. The mesh refinement simulation used inhomogeneous particle loading and loaded 1 ion particle per cell regardless of the cell size and the uniform-grid simulation loads 1 particle per cell. In Section~\ref{sec:positron}, the positron simulation utilized a grid size of $512 \times 512 \times 1024$ cells for the coarsest resolution, with a refinement ratio of 8. In this case, the speedup achieved with mesh refinement was approximately 80 compared to the fine resolution everywhere simulation. It is worth noting that the speed-up in this scenario is relatively modest due to the smaller refinement ratio. As noted above, we fixed the particles per cell regardless of the cell size. Thus, a lot of the particles are clustered in a few MPI partitions and this can cause load imbalance. Further optimization of particle operation load balancing could improve computational efficiency in this context. 

\section{Conclusion}

We describe a mesh refinement algorithm developed to accelerate full 3D QuickPIC simulations of two bunch PWFA simulations. The code applies high-resolution refined meshes in specific regions, typically surrounding the trailing beam in two bunch PWFA simulations, and the grid sizes of those meshes progressively change to coarser resolutions in the remaining parts of the simulation domain. Currently, the algorithm uses square cells in the two transverse directions. The cross section of the refined domains also needs to be rectangular but this is not a severe limitation. A fast multigrid Poisson solver has been implemented to solve the refined meshes, and the solver has been parallelized with both MPI and OpenMP. The coarsest mesh can be solved by both FFT and multigrid Poisson solvers. The code also enables inhomogeneous particle loading, where we can initialize more plasma particles inside the refined meshes. We resolve unbalanced computational load of the field solver by spreading fine grids that are usually clustered in one location into all processors, and gathering them back after solving the Poisson equation on refined meshes. The scalability of the code has also been improved significantly by using pipelining. We also developed a preliminary adaptive mesh refinement by choosing certain refined mesh domains and grid sizes during the simulation to save the computational time for simulations with an evolving beam size. Several methods have been used to benchmark the mesh refinement algorithm. The evolution of both symmetric and asymmetric witness beams have been simulated with ion motion. For the symmetric electron witness beam, the emittance growth simulated by mesh refinement is consistent with the quasi-3D QPAD simulations with fine resolution throughout the entire simulation domain. We have demonstrated that the use of multiple levels of refinement greatly reduces spurious fields near the boundaries between different levels of refinement. Additionally, we show that the mesh refinement code can be applied to simulate PWFA with high degrees of azimuthal asymmetry. For the symmetric witness positron beam, a single time step simulation with mesh refinement agrees with fully resolved 3D QuickPIC. An example of adaptive mesh refinement has also been presented and the results agree with QPAD. On problems of interests, speed-ups between $10\sim100$ has been found.
	
\section*{Acknowledgements}
This worked was supported by the U.S. Department of Energy (DOE), Offices of Advanced Scientific Computing Research and  of High Energy Physics (HEP) through the Scientific Discovery through Advanced Computing (SciDAC) project, ComPASS-4,  under Contract Nos.\ DE-AC02-06CH11357 and DE-AC02-05CH11231, and a FNAL subcontract 644405, and by a DOE HEP grant DE SC0010064, and by the U.S. National Science Foundation through award \# 2108970. We also acknowledge useful conversations with Dr. Carlo Benedetti.
\appendix
\section{}\label{sec:appendix}
Multigrid methods~\citep{briggs2000multigrid}  have emerged as powerful techniques for efficiently solving differential equations within arbitrary spatial domains and for complex boundary conditions. Here we describe in detail the multigrid algorithm we use to solve for the fields on a refined mesh. This multigrid solver employs the V-cycle approach, which combines restriction, interpolation and smoothing operations. Recall that previously we defined $\Omega_{l, r}$ as the domain for the refinement grid at level $l$ with refinement ratio $2^r$ between the current level $l$ and the next coarser level, $l-1$. To solve the Poisson like equations on $\Omega_{l, r}$, a hierarchy of grids is established by iteratively reducing the grid points in both the $x$ and $y$ directions by factors of 2. This set of meshes is denoted as $\Omega_{l,m}$. For refinement level $l$, where $M_l$ meshes are created for the multigrid algorithm, we denote these meshes as $\Omega_{l, r}$, $\Omega_{l,r-1}$, ..., $\Omega_{l,r - M_l + 1}$ and the grid sizes of these meshes varies from $h_l$ to $2^{M_l-1} h_l$. We will establish such a hierarchy of grids for the field solution $F_{l, m}$, correction $E_{l, m}$, and residual $R_{l,m}$, the definition of which will be explained when we discuss the multigrid algorithm. As an example, consider the $l = 2$ grid with $r = 1$, so the refinement ratio between level 2 and level 1 is 2. For simplicity, we assume this mesh is square, and the global indices ranging from 252 to 264 are created with resolution $h_{2,1}=h/2$, denoted as $\Omega_{2,1}
$, and $h$ is the resolution of $l=1$. Subsequently, two additional meshes $\Omega_{2,0}$ and $\Omega_{2,-1}$, where the fields on these meshes are denoted as $F_{2, 0}$ and $F_{2,-1}$, and lower resolutions, $h_{2,0} = 2h_{2,1}$ and $h_{2,-1} = 4h_{2,1} $,  respectively are generated. The indices span from 126 to 132 for $F_{2, 0}$ and 63 to 66 for $F_{2, -1}$. This recursive process leads to a pyramid of grids sharing the same boundary, each representing a progressively coarser level of resolution. 

\begin{algorithm}[!h]
\caption{Multigrid cycle}\label{alg:multigrid}
\begin{algorithmic}[1]
    \Function{multigrid\_solve}{$F_l,S_l$}	
    
    \State Solve $l = 1$ field from $\nabla_{\perp}^2 F_1 = S_1$
    \State Initial value of $F_2 = F^0_{2,r}$ interpolated from $F_1$
    \State Calculate residual $R^1_{2,r} = S_2 - \nabla_{\perp}^2 F^0_{2,r}$
    \For{Vcycle iteration $n = 1, 2 ....$}   \Comment{V-cycle}
    \For {grids $\Omega_{2,m}$ on level 2 from $\Omega_{2,r}$, $\Omega_{2,r-1}$ ... to $\Omega_{2,r - M_l + 1}$}  \Comment{Downward pass}
    \State Restrict residual from finer mesh $R^n_{2,m}$ to coarse $R^n_{2,m-1}$
    \EndFor
    \State Solve $\nabla_{\perp}^2 E^n_{2, r - M_l + 1} = R^n_{2,r-M_l+1}$ with the conjugate gradient solver
    \For {grids $\Omega_{2,m}$ from $\Omega_{2,r -M_l + 2}$, ... to $\Omega_{2,r}$}  \Comment{upward pass}
    \State Interpolate correction $E^n_{2, m-1}$ to $E^n_{2, m }$ 
    \State Smooth field $E^n_{2, m}$  with the red-black GS algorithm
    \EndFor
    \State Add correction to the solved field $ F^{n }_{2,r} = F^{n-1}_{2,r} +  E^{n}_{2,r}$
    \State Calculate residual $R^{n+1}_{2,r} = S_2 - \nabla_{\perp}^2 F^n_{2,r}$
    \If {$ \lVert R^{n+1}_{2,r}  \rVert < \delta$ }
    \State \textbf{Break}
    \EndIf

    \EndFor
    
    \EndFunction
    
\end{algorithmic}
\end{algorithm}

As an example, consider solving Poisson's equation $\nabla_{\perp}^2 F_l = S_l$, and assume there are only two levels of refinement, $l$ =  1 and 2, where the second level has a refinement ratio of $2^{r}$, and the mesh is denoted as $\Omega_{2, r}$. Recall that the source terms, $S_1$ and $S_2$ have been obtained from the particle deposition. The workflow of the multigrid V cycle for the $l = 2$ mesh is shown as Algorithm \ref{alg:multigrid}. After solving for the coarsest,  $l = 1$ mesh $F_1 = F_{1, 1}$ with an FFT solver, the initial field $F^{n = 0}_{2,r}$, including the boundary values, is interpolated from $F_{1, 1}$ by the bilinear interpolation scheme. We use subscript $n$ to represent current V-cycle number.  $M_2$ meshes are created for the multigrid algorithm for the correction $E_{2, m}$, residual $R_{2,m}$ and fields $F_{2, m}$ separately. We next start the V-cycle multigrid algorithm to solve $\nabla_{\perp}^2 F_2 = S_2$ using the initial value of $F^0_{2,r}$ interpolated from the $F_1$ solution. For the $n=1$ part of the V-cycle, the residual on $\Omega_{2,r}$ is calculated by $R^{n=1}_{2,r} \equiv S_2 - \nabla_{\perp}^2 F^0_{2,r}$, and this is then recursively been passed down (restricted) to coarser levels $\Omega_{2,r-1}$,...,$\Omega_{2,m}$, ... $\Omega_{2,r-M_2 + 1}$ with resolution ($2h_{2,r}$,... , $2^{r - m}h_{2,r}$..., $2^{M_2 - 1}h_{2,r}$). A technique called full weighting~\citep{briggs2000multigrid} is used to restrict the residual  $R^n_{2,m}$ from finer grid levels $\Omega_{2, m}$ to coarser grid levels $\Omega_{2, m-1}$.  This full weighting scheme effectively transfers information from finer grids to coarser grids and can be represented by the following equation:

\begin{equation*}
\begin{aligned}
 R^n_{2,m-1}(i,j) = 1/4 \left[R^n_{2, m}(2i,2j)\right] + 1/8 \left[  R^n_{2, m}(2i+1, 2j) + R^n_{2, m}(2i-1, 2j) \right.\\
\left. +  R^n_{2,m}(2i, 2j + 1) + R^n_{2,m}(2i, 2j - 1)\right] \\
 + 1/16\left[  R^n_{2,m}(2i+1, 2j+1) +  R^n_{2,m}(2i-1, 2j+1) + \right.\\
 \left.  R^n_{2,m}(2i+1, 2j-1) + R^n_{2,m}(2i - 1, 2j - 1)\right].
 \end{aligned}
 \end{equation*}

Here, the values of \(R^n_{2,m-1}(i,j)\) on grid indexes of $(i,j)$ are calculated based on the values of neighboring points on the finer grid (\(R^n_{2,m}(2i,2j)\)). Note we are using global indices to label grid points so that $(i,j)$ on mesh $\Omega_{2,m-1}$ is at the same position as $(2i, 2j)$ on mesh $\Omega_{2,m}$. The coefficients in the equation determine the weighting of these grid points during the restriction process.

After the restriction of residual to the mesh $\Omega_{2,r-M_2+1}$ with coarsest resolution $2^{M_2-1}h_{2,r}$, a conjugate gradient solver is implemented to solve the correction $E^n_{2,r-M_2 + 1}$ to the field $F^n_{2,r-M_2 + 1}$ from equation $\nabla_{\perp}^2 E^n_{2, r - M_2 + 1} = R^n_{2,r-M_2+1}$.   Subsequently, in the upward pass of the V-cycle, the coarser correction $E^n_{2, m - 1}$ is passed to the finer grid $E^n_{2, m}$ through interpolation.  After each interpolation to a finer grid $\Omega_{2, m}$, the red-black Gauss-Seidel (GS) algorithm is applied to smooth the correction $E^n_{2, m}$ on each subgrid $\Omega_{2, m}$. This smoothing operation involves iteratively updating $E^n_{2, m}$ at each grid point using its neighboring points. The equations for the red-black Gauss-Seidel algorithm~\citep{briggs2000multigrid} to smooth $\nabla_{\perp}^2 E^n_{2,m} = \prescript{}{}R_{2,m}$ are expressed as follows:
\begin{align*}
\prescript{k+1}{} E^n_{2, m}(i,j) = \frac{1}{4}\left[\prescript{k}{} E^n_{2, m}(i-1, j)+ \prescript{k}{}E^n_{2, m}(i+1, j)+ \prescript{k}{}E^n_{2, m}(i, j-1) \right. \\
\left. +\prescript{k}{}E^n_{2, m}(i, j+1) - h_{2, m}^{2} R^n_{2, m}(i, j)\right].
\end{align*}

$R^n_{2,m}(i, j)$ is the residual term at grid point $(i, j)$ that we obtained in the downward pass and $\prescript{k+1}{} E^n_{2,m}(i, j)$ represents the updated field values at grid point $(i, j)$ after the $k$-th iteration of the red-black Gauss-Seidel algorithm. For parallelization, we alternate between red points and black points to update $\prescript{k+1}{} E^n_{2,m}$~\citep{briggs2000multigrid}. Typically, for each grid $\Omega_{2,m}$ except $\Omega_{2,r-M+1}$, which has been solved by the conjugate gradient algorithm, we choose iteration numbers $k$ around 2. The correction  $E^{n = 1}_{2,r}$ obtained at the end of the first V-cycle  will be added to the field $F^{n = 1}_{2,r} = F^{n=0}_{2,r} +  E^{n = 1}_{2,r}$ . Then the new residual $R^{n = 2}_{2,r}$  of the next V-cycle will be calculated from the updated field by $R^{n=2}_{2,r} \equiv S_2 - \nabla_{\perp}^2 F^1_{2,r}$. The  V-cycle loop will be repeated until norm of the residual $\| R^{n}_{2,r} \|$ is less than a preset value $\delta$ (usually set close to the machine precision).

\bibliographystyle{elsarticle-num-names} 
\bibliography{ref}
\end{document}